\documentclass[journal=jctcce,manuscript=article]{achemso}

\usepackage{pdfpages}
\usepackage[version=3]{mhchem} 
\usepackage{amsmath,amssymb,amsthm, amsfonts, mathtools, dsfont} 
\usepackage{bbm,bm,tensor, braket}
\usepackage{eqnarray,array,enumerate}
\usepackage{csquotes}
\usepackage{siunitx}
\usepackage{booktabs}
\usepackage{graphicx,wrapfig, caption, float, subcaption, epstopdf, setspace}
\usepackage{hyperref}
\usepackage{longtable}
\usepackage[section]{placeins}
\usepackage{multicol, multirow}
\usepackage{tikz,tkz-euclide}
\usetikzlibrary{shapes.geometric,arrows,arrows.meta, calc, positioning, automata, shadows, backgrounds,decorations.markings,decorations.pathreplacing}
\usepackage{tkz-fct}
\usetikzlibrary{intersections}
\usepackage{pgfplots}
\pgfplotsset{compat=1.9}
\usepgfplotslibrary{fillbetween}
\usepackage{pagecolor}\usepackage{amssymb}
\usepackage{algorithm, algpseudocode}
\usepackage{epstopdf}

\usepackage{cleveref}

\newcommand*{\citen}[1]{%
  \begingroup
    \romannumeral-`\x 
    \setcitestyle{numbers}%
    \cite{#1}%
  \endgroup   
}

\algrenewcommand\algorithmiccomment[1]{\hfill #1}

\def\br{\ensuremath\bm{r}}

\author{Arno F{\"o}rster}
\email{a.t.l.foerster@vu.nl}
\affiliation{Theoretical Chemistry, Vrije Universiteit, De Boelelaan 1083, NL-1081 HV, Amsterdam, The Netherlands}
\author{Lucas Visscher}
\affiliation{Theoretical Chemistry, Vrije Universiteit, De Boelelaan 1083, NL-1081 HV, Amsterdam, The Netherlands}

\title{Quasiparticle Self-Consistent $GW$-Bethe-Salpeter equation calculations for large chromophoric systems}


\DeclareUnicodeCharacter{2212}{-}
\begin{document}

\begin{abstract}
The $GW$-Bethe-Salpeter Equation (BSE) method is promising for calculating the low-lying excited states of molecular systems. However, so far it has only been applied to rather small molecules, and in the commonly implemented diagonal approximations to the electronic self-energy it depends on a mean-field starting point. We describe here an implementation of the self-consistent and starting-point independent quasiparticle self-consistent (qs$GW$)-BSE approach which is suitable for calculations on large molecules. We herein show that eigenvalue-only self-consistency leads to an unfaithful description of certain excitonic states for Chlorophyll dimers while the qs$GW$-BSE vertical excitation energies (VEE) are in excellent agreement with spectroscopic experiments for Chlorophyll monomers and dimers measured in the gas phase. On the other hand, VEEs from time-dependent density functional theory calculations tend to disagree with experimental values and using different range-separated hybrid (RSH) kernels changes the VEEs by up to 0.5 eV. We use the new qs$GW$-BSE implementation to calculate the lowest excitation energies of the six chromophores of the photosystem II (PSII) reaction center (RC) with nearly 2000 correlated electrons. Using more than 11000 (6000) basis functions, the calculation could be completed in less than 5 (2) days one a single modern compute node.
In agreement with previous TD-DFT calculations using RSH kernels on models that do also not include environment effects, our qs$GW$-BSE calculations only yield states with local character in the low-energy spectrum of the hexameric complex. Earlier work with RSH kernels has demonstrated that the protein environment facilitates the experimentally observed interchromophoric charge transfer. Therefore, future research will need to combine correlation effects beyond TD-DFT with an explicit treatment of environment electrostatics. 

\end{abstract}

\section{\label{sec:introduction}Introduction}
\begin{figure}[hbt!]
    \centering
    \includegraphics[width=1.0\textwidth]{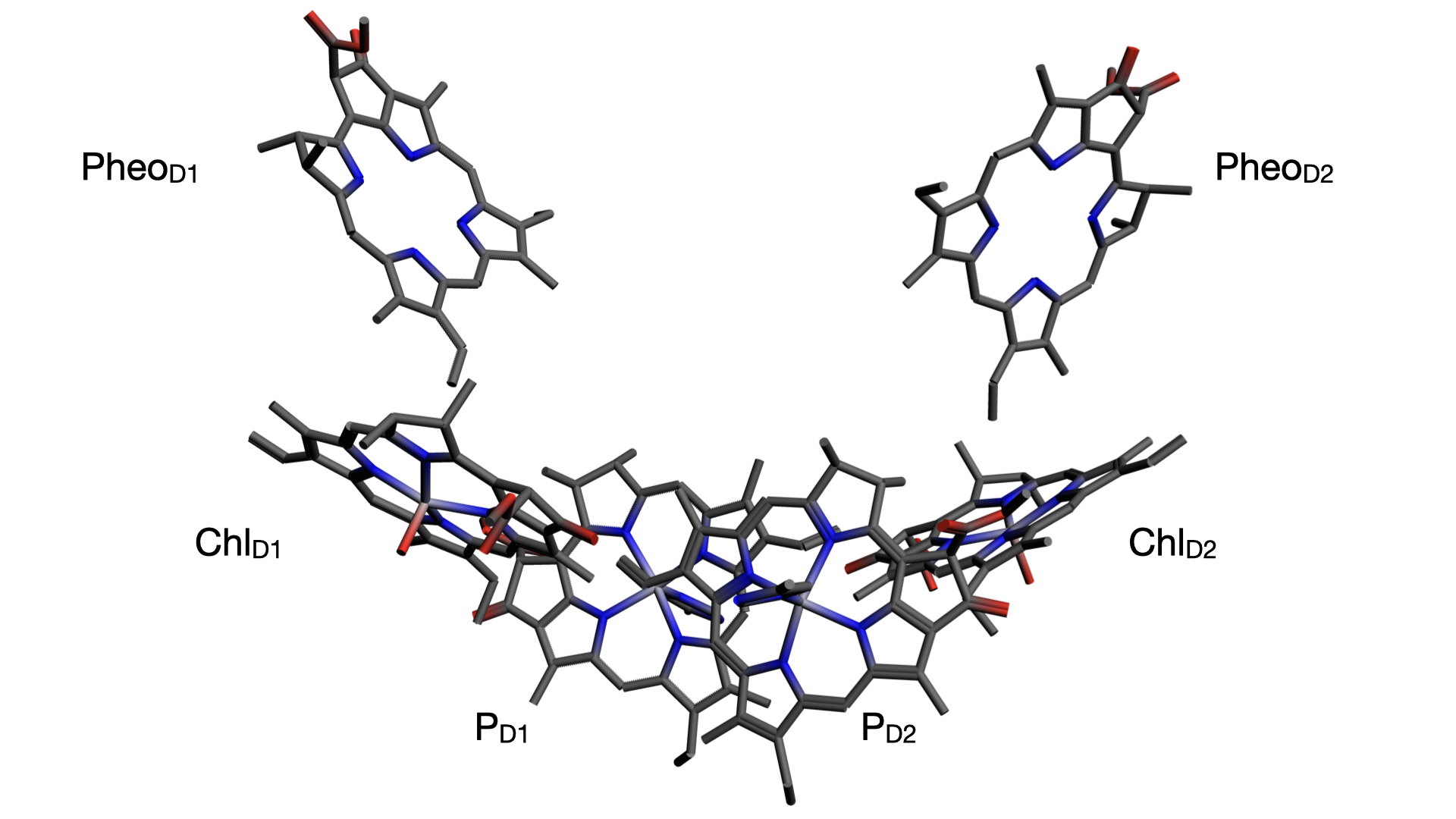}
    \caption{Chromophores of the Photosystem II reaction center.}
    \label{fig::chromophores}
\end{figure}

\protect
The absorption of photons by a molecule or a material upon interaction with electric radiation is a key process in the conversion of light into chemical or electrical energy. In the photosystem II (PSII) reaction center (RC), photons are captured by chromophoric complexes which then leads to the generation of free charge carriers.\cite{Mirkovic2017} In the first step of this process an electron-hole pair is formed, where electron and hole are bound due to their Coulombic interaction.\cite{Croce2020} Such bound electron-hole states are commonly referred to as excitons and correspond to the energies of the absorbed photons.\cite{Kasha1965} 
In the current work we look at the characterization of such low-lying excited states of the RC of PSII which is at the heart of photosynthetic function.\cite{Reimers2016} As shown in figure~\ref{fig::chromophores}, the PSII RC contains six chromophores, a "special pair",\cite{Yin2007, Renger2010} of two Chlorophyll $a$ (chla) molecules (P\textsubscript{D1} and P\textsubscript{D2}), flanked by two more chla (Chl\textsubscript{D1} and Chl\textsubscript{D2}) and two Pheophytin $a$ (Pheo\textsubscript{D1} and Pheo\textsubscript{D2}) molecules, with around 2000 electrons in total. By now, it has been firmly established that the primary events of charge separation in PSII are determined by a complex interplay of all these six chromophores.\cite{Sirohiwal2020a} Therefore, all six chromophores should ideally be treated on a quantum mechanical level and their couplings need to be taken into account.

In most current calculations of larger biomolecular complexes, one resorts to Hartree-Fock (HF)\cite{Kitagawa2011,Polyakov2018} or Time-dependent (TD) Density Functional Theory (DFT) with a range-separated hybrid (RSH) exchange-correlation kernel\cite{Reimers2013, Frankcombe2015, Kavanagh2020, Sirohiwal2020a, Sen2021, Tamura2021, Sirohiwal2022}. RSHs frequently offer good agreement with experiment for Chla monomers and dimers,\cite{Lopez-Tarifa2017, Anda2019, Sen2021} but large deviations with respect to more advanced multi-configurational\cite{Cai2006,Anda2019} and wave-function based methods have also been observed.\cite{Sirohiwal2021} To mitigate such errors, RSHs can be parametrized empirically for each system under investigation (as for example done in references in \citen{Saito2018} and \citen{Shao2020}), but this makes them non-transferable and unreliable for general applications. Systematic tuning procedures for range-separated functionals have been suggested as well.\cite{Stein2010, Refaely-Abramson2011, Refaely-Abramson2012, Kronik2012} Those however always require to perform exploratory calculations to find the ideal range-separation parameter. Furthermore, heterogeneous systems like multi-chromophoric complexes might require different range separation parameters for different regions of the complex.\cite{Karolewski2013}

Turning to wave-function based methods for excited states, we find the second-order algebraic diagrammatic construction scheme (ADC(2))\cite{Schirmer1982, Schirmer1983} and coupled cluster\cite{Coester1958, Coester1960, Cizek1966, Cizek1969, Paldus1972} with approximate doubles (CC2)\cite{Christiansen1995} easy to apply and reasonably cost-efficient. CC2 results are typically in good agreement with more involved methods like equation-of-motion (EOM) CC\cite{Stanton1993} with singles and doubles (EOM-CCSD) or similarity-transformed (ST) EOM\cite{Nooijen1997,Nooijen1997a}-CCSD\cite{Loos2020e, Monino2021}. For these methods we are aware of one study of a tetrameric model by Suomivuori et al.\cite{Suomivuori2016} using ADC(2) together with the spin-opposite-scaled\cite{Winter2011} and reduced-virtual-space (RVS)\cite{Send2011} approximations. Unfortunately, they did not include the Pheophytin chromophores in their calculations, which are known to play a key role in the initial charge separation immediately after photoexcitation.\cite{Groot2005, Romero2012, Tamura2021, Yoneda2022} This is potentially possible, but we note that most applications of wave-function based methods\cite{Anda2019, Sirohiwal2020,Sirohiwal2021,Sirohiwal2021a} focus on single chromophores. Utilizing subsystem methods\cite{Wesolowski1993, Neugebauer2005, Gomes2012, Manby2012, Hofener2012, Raghavachari2015, Tolle2021} the applicability of these methods can be extended. In this family of methods one describes the full RC by an effective Hamiltonian with a limited amount of levels for each chromophore. The information needed to build such an effective Hamiltonian are the monomeric  excitation energies as well as the inter-monomeric couplings. These parameters can be computed in a first principles manner with various electronic structure methods\cite{Neugebauer2007, Hofener2016, Leng2019}. While the subsystem approach can be used with high-level monomer calculations, a drawback is that commonly used approximations to calculate the couplings between the chromophores are often not accurate enough.\cite{Frahmcke2006, Send2011, Lopez-Tarifa2017} In the current work we will therefore examine how large a system can be treated directly without having to resort to partitioning and subsystem methods. As the states of interest are the lowest energy ones, we thereby focus on a limited number of states, but describe them in a supermolecular fashion that fully accounts for all intermolecular couplings of the chromophores.

Our approach is based on the $GW$-BSE method that we will briefly summarize in the following. We first note that energy levels of the excitonic states correspond to the poles of the 2-particle generalized susceptibility.\cite{Sham1966a, Hanke1979, Hanke1980, Strinati1988} This quantity can be obtained from the interacting single-particle Green's function $G_1$ and the electronic self-energy $\Sigma$, a non-local, non-Hermitian, and frequency dependent one-electron operator, via a Bethe-Salpeter equation (BSE)\cite{Salpeter1951, Gell-Mann1951, Kugler2018}. $G_1$ is obtained from a Dyson equation with $\Sigma$ as its kernel, while $\Sigma$ itself depends implicitly on the 2-particle Green's function.\cite{martin2016, Kugler2018, Rohringer2018}
As obtaining the full generalized susceptibility requires $N^6$ operations, it is advantageous to decouple the BSE from the Dyson equation for $G_1$. This is done by using an approximation to the self-energy which only depends on the density-density response\cite{Petersilka1996, Onida2002}. A popular example is the $GW$ approximation (GWA), with the screened Coulomb interaction $W$\cite{Hubbard1957, Phillips1961} calculated within the random phase approximation (RPA).\cite{Hedin1965} Typically, the Dyson equation for $G_1$ is solved within the GWA first. Only afterwards, the non-interacting 2-particle Green's function and the corresponding kernel in its zero-frequency limit are constructed and one solves for a few or all roots of the generalized susceptibility.\cite{Onida1995,Rohlfing1998,Rohlfing2000} If only a few excitonic states are needed, one may thereby use computationally efficient iterative diagonalization techniques\cite{Stratmann1998, Rohlfing2000}. This procedure is known as the $GW$-BSE method and is increasingly applied to compute the lowest electronically excited states of molecular systems.\cite{Blase2011a,
Baumeier2012,
Duchemin2012, 
Faber2013,
Baumeier2014,
Korbel2014,
Boulanger2014,
Jacquemin2015,
Blase2016,
Ziaei2016,
Hung2017,
Krause2017,
Rangel2017,
Gui2018,
Duchemin2018,
Blase2018,
Holzer2018,
Sharifzadeh2018,
Holzer2019,
Leng2019,
Tirimbo2020,
Liu2020,
Loos2020,
Wang2020,
Kehry2020,
Patterson2020,
Grupe2021,
Holzer2021,
Hashemi2021,
Tolle2021,
DeQueiroz2021,
Yao2022}

For such applications, the $GW$ part is typically the computational bottleneck of a $GW$-BSE calculation.\cite{Gui2018, Blase2018, Hashemi2021} The issue has been addressed over the last years: Many implementations of $G_0W_0$ and ev$GW$ with reduced asymptotic scaling with system size have been developed\cite{Wilhelm2018, Fujita2019, Kim2020, Forster2020b, Wilhelm2021, Duchemin2021a, Fujita2021, Neuhauser2014, Vlcek2017, Vlcek2018, Weng2021} often producing results in excellent agreement with conventional $GW$ implementations.\cite{Wilhelm2018, Wilhelm2021, Duchemin2021a} Another issue is related to the common approximations in solving the $GW$ equations. Typical calculations start from a Kohn--Sham (KS)-DFT or HF Green's function followed by a perturbative update of the QP energies ($G_0W_0$).\cite{Hybertsen1985, Hybertsen1986} This procedure comes with the notable disadvantage that the outcome of such a calculation will heavily depend on the choice of the underlying exchange-correlation (XC) functional.\cite{Bruneval2013, Bruneval2015, Jacquemin2015, Knight2016, Caruso2016} Achieving self-consistency in the eigenvalues only (ev$GW$) can remove this dependence on the initial density functional approximation to a large extent but not completely.\cite{Gui2018, Loos2021, Hashemi2021}

Instead, one can also start from the full $GW$ self-energy and take the Hermitian part only to arrive at a set of effective single-particle equations.\cite{Huser2013, Nakashima2021} In QP self-consistent $GW$ (qs$GW$), then only the low-frequency limit of the self-energy is considered\cite{Faleev2004, VanSchilfgaarde2006, Kotani2007}, and the non-interacting $G_1$ closest to the $GW$ $G_1$ is selected.\cite{Ismail-Beigi2017} 
While this approach has been shown to be more accurate than $G_0W_0$ and ev$GW$ for a wide range of molecular systems\cite{Forster2022},  qs$GW$ has until now rarely been used in molecular calculations. With only a few exceptions,\cite{Kutepov2020,Forster2021a} low-order scaling $GW$ algorithms only target the screened Coulomb interaction, since this requires only evaluation of the diagonal elements of the self-energy. The computational cost for obtaining the full self energy is much larger, and most implementations therefore become inefficient if the full self-energy is required.
To address this issue, we have recently presented a low-order scaling implementation of qs$GW$.\cite{Forster2021a} In the present work, we combine it with an efficient solver for the BSE, resulting in a fast, low-scaling, and starting-point independent implementation of the GW-BSE approach. 

The $GW$-BSE method has recently been shown to reproduce experimental low-lying excitation energies of Chls with high accuracy.\cite{Hashemi2021, Li2022} So far, it has only been applied to monomeric models of PSII. In this work, we will first give a brief account of the (low-scaling) implementation of the $GW$-BSE approach in section~\ref{sec:theory}. After describing some technical details of our calculations in section~\ref{sec:computational_details}, in section~\ref{sec:results} we first contrast qs$GW$-BSE to ev$GW$-BSE for single chromophores and chromophore dimers and confirm the excellent agreement of the former with experiment. We then use the qs$GW$-BSE implementation to calculate the low-lying excitation of the hexameric complex with 2000 correlated electrons in total. Finally, section~\ref{sec:conclusion} summarizes and concludes this work.

\section{\label{sec:theory}Theory}\protect
\subsection{The $GW$-BSE formalism}
The interacting $n$-particle Green's functions corresponding to an $N$-electron system with ground state $\Psi^{(N)}_0$ are defined by 
\begin{equation}
    G_n(1, \dots 2n) = (-i)^n
    \left\langle
    \Psi^{(N)}_0
    \Big|
    \mathcal{T} 
    \left[
    \hat{\psi}^{\dagger}(1) 
    \hat{\psi}(2) 
    \dots 
    \hat{\psi}^{\dagger}(2n-1) 
    \hat{\psi}(2n) 
    \right]
    \Big| 
    \Psi^{(N)}_0
    \right\rangle \;.
\end{equation}
Here, $\mathcal{T}$ is the time-ordering operator, $\hat{\psi}$ is the field operator and a number $1 = (\br_1, \sigma_1, t_1)$ collects space, spin-and time indices. The relevant cases are $n =1, 2$. For the $n=2$ case, we further restrict ourselves to the excitonic part only with $t_3 = t_4$ and $t_1 = t_2$.

The single-particle Green's function can be related to its non-interacting counterpart $G^{(0)}_1$ by a Dyson equation
\begin{equation}
\label{Dyson}
  G^{(0)}_1(1,2) = G^{(0)}_1(1,2) + G^{(0)}_1(1,3)\Sigma(3,4)G_1(4,2) \;,
\end{equation}
in which the self-energy operator $\Sigma$ appears.\cite{Luttinger1960a} In \eqref{Dyson} and in the following, integration over repeated indices is implied. 
The reduced 2-particle Green's function
\begin{equation}
    L(1,2,3,4) = -G_2(1,2,3,4) + G_1(1,2)G_1(3,4) \;,
\end{equation}
fulfills a BSE,\cite{Strinati1988, Romaniello2009b}
\begin{equation}
\label{BSE}
  L(1,2,3,4) = L^{(0)}(1,2,3,4) + L^{(0)}(1,6,2,5)
  \frac{\delta \Sigma(5,7)}{\delta G_1(8,6)}L(8,2,7,4) \;,
\end{equation}
where\cite{Romaniello2009b}
\begin{equation}
    L^{(0)}(1,2,3,4) = G_1(1,4)G_1(2,3) \;.
\end{equation}
The local Hartree kernel is obtained by approximating $\Sigma$ with the Hartree potential,
\begin{equation}
    \Sigma_H(1,2) = v_H(1)\delta(1,2)
    = -i \delta(1,2)\int d3\; v_c(1,3)G_1(3,3^+) \;,
\end{equation}
where $v_c$ is the Coulomb potential and $1^+ = \lim_{\eta \rightarrow 0^+}(\br_1, \sigma_1, t_1 + \eta)$.
Calculating 
\begin{equation}
\label{vHartree}
    \frac{\delta }{\delta G_1(4,2)} \int d3\; v_c(1,3)G_1(3,3^+) = v_c(1,2)\delta(3,4)\delta(3,2) \;,
\end{equation}
 and inserting the result into \eqref{BSE} one then obtains 
\begin{equation}
\label{P}
    P(1,2) = P^{(0)}(1,2) + P^{(0)}(1,3)v_c(3,4)P(4,2) \;,
\end{equation}
  with 
\begin{equation}
\label{PL}
P(1,2) = L(1,2^{+},1^{+},2)
\end{equation}
being the $v_c$-reducible density-density response function in the RPA and
\begin{equation}
\label{pdef}
    P^{(0)}(1,2) = -i G(1,2)G(2,1^{+}) \;.
\end{equation}
 $P$ is related to the screened Coulomb interaction $W$ by\cite{Hubbard1957}
\begin{equation}
\label{W}
    W(1,2) = v_c(1,2) + v_c(1,3) P (3,4)v_c(4,2) \;,
\end{equation}
which can be used to define the $GW$ self-energy, 
\begin{equation}
\label{sigma}
    \Sigma^{(GW)}(1,2) = v_H(1,2)  + i G(1,2)W(1^+,2) \;.
\end{equation}
\Cref{Dyson,P,pdef,W,sigma} constitute a self-consistent set of equations, usually referred to as the $GW$-approximation. 

By splitting the self-energy into Hermitian and anti-Hermitian part and discarding the latter one, we can restrict the solution of \eqref{Dyson} to its QP part only.\cite{Layzer1963, Sham1966, Huser2013, Nakashima2021} We then have an effective single-particle problem and restricting the self-energy further to its static limit and transforming to the molecular orbital basis $\left\{\phi_n\right\}_{n = 1 \dots N}$ (in which the single-particle Hamiltonian is diagonal), we arrive at
\begin{equation}
\label{DysonqsGW}
    \sum_{m}\left\{\left(\epsilon_n - \epsilon^{QP}_n\right)\delta_{nm} + \frac{1}{2}\left[\Sigma_{nm}^{(GW)}(\epsilon_n)
    + \Sigma_{mn}^{(GW)^{*}}(\epsilon_n)
    \right]\right\}\phi_n = 0 \;,
\end{equation}
where the $\epsilon_n$ are the single-particle energies. Solving \cref{DysonqsGW,P,pdef,W,sigma} self-consistently is known as the qs$GW$ approximation within the RPA.\cite{Faleev2004, VanSchilfgaarde2006, Kotani2007}

After solving  the qs$GW$ equations self-consistently, we can then use the zero-frequency limit of the self-energy \eqref{sigma} for the kernel of \eqref{BSE}. As it is typically done, we also set $\frac{\delta W}{\delta G} \approx 0$. This is referred to as the qs$GW$-BSE approach. After Laplace transformation to the complex frequency plane, eq.~\eqref{BSE} can be transformed into an eigenproblem in a basis of particle-hole states whose solution provides the Lehmann representation of $L$ (see for example ref.~\citen{Maggio2016} or ref.~\citen{Sander2015} for detailed derivations),
\begin{equation}
\label{casida}
    \begin{pmatrix}
    \mathbf{A} & \mathbf{B} \\ 
    -\mathbf{B} & -\mathbf{A} \\ 
    \end{pmatrix}
    \begin{pmatrix}
    \mathbf{X} \\
    \mathbf{Y}\\
    \end{pmatrix}_{S}
    = 
    \Omega_s
     \begin{pmatrix}
    \mathbf{X} \\
    \mathbf{Y}\\
    \end{pmatrix}_{S} \;.
\end{equation}
$\Omega_S$ is a neutral excitation energy, $(\mathbf{X}, \mathbf{Y})^T_{S}$ contains the expansion coefficients of the corresponding eigenvector and for a closed-shell system the matrix elements of $\mathbf{A}$ and $\mathbf{B}$ are respectively defined as
\begin{equation}
\label{matrix}
\begin{aligned}
    A_{ia,jb} = & 
    2v_{c_{iajb}} - W(\omega= 0)_{ijab} + \delta_{ab} \delta_{ij} 
    \left(\epsilon^{QP}_i - \epsilon^{QP}_a\right)  \\ 
    B_{ia,jb} = & 
    2v_{c_{iajb}} - W(\omega= 0)_{ajbi} \;,
\end{aligned}
\end{equation}
where we have chosen to reserve the labels $i,j, \dots$ for occupied and $a,b,\dots$ for virtual orbitals. The QP energies entering the equations are the ones from \eqref{DysonqsGW}. 

\subsection{Implementation}
For our implementation of the qs$GW$ methods we refer to our previous work.\cite{Forster2020b, Forster2021, Forster2021a} We expand single-particle Green's functions and the self-energy in a basis of Slater type functions (primary basis) which is related to the MOs by
\begin{equation}
\label{primary-basis}
  \phi_i(\br) = \sum_{\mu} c_{i\mu}\chi_{\mu} (\br) \;,  
\end{equation}
while all quantities appearing in \eqref{W} are expanded in a basis of auxiliary fit functions (auxiliary basis). We then switch to the particle-hole basis to solve \eqref{casida}, whereby the matrix elements in \eqref{matrix} are expanded in the basis of MOs.

Since we do not use the screened interaction at zero frequency in our $GW$ implementation, we calculate the zero-frequency component of $P$ from the imaginary time representation of the polarizability by 
\begin{equation}
    P(\omega = 0) = \frac{1}{2\pi} \int P(i \tau) d\tau \;,
\end{equation}
and we then use \eqref{W} to obtain $W(\omega=0)$. 

 Replacing the matrix elements of the screened Coulomb interaction by the ones of the bare one in \eqref{matrix}, and using the HF self-energy in \eqref{DysonqsGW}, the TD-HF method is obtained. It is clear, that any solver which can be used to solve \eqref{casida} in the TD-HF case, can also be used for $GW$-BSE. We use an extension of the Davidson algorithm\cite{Davidson1975} originally proposed by Stratmann and Scuseria.\cite{Stratmann1998} It solves \eqref{BSE} by projecting the generalized problem
\begin{equation}
\label{generalized}
    \left(\mathbf{A}-\mathbf{B}\right)
    \left(\mathbf{A}+\mathbf{B}\right)
    \left(\mathbf{X}+\mathbf{Y}\right)
    = \Omega^2_S \left(\mathbf{X}+\mathbf{Y}\right) \;, 
\end{equation}
on a sequence of orhonormal subspaces
\begin{equation}
\label{subspaces}
\mathrm{span}\left\{b^{(n)}_1, \dots b^{(n)}_{k}\right\} \;,
\end{equation}
in which \eqref{generalized} is solved. $k$ denotes the size of the $n$th subspace and the $b_k$ are linear combinations of particle-hole states. The vectors forming the subspace are then updated until the subspaces are converged. The procedure can be interpreted as an iterative optimization of the basis of particle-hole states, where the part which does not carry useful information (i.e. the particle-hole transitions which do not contribute to the low-lying excitons) is projected out. 

The time-determining step in the diagonalization is the projection of the eigenproblem in the full space on the subspaces. The term containing the bare Coulomb potential is easily evaluated following the procedure in \citen{VanGisbergen1999}. For the matrix elements of the screened interaction in the $(n+1)$th subspace iteration, we define a column in the subspace labeled by $s_i, s_j, \dots$, $s_a, s_b, \dots$, respectively, as
\begin{equation}
\label{davidson-matrix}
    (\mathbf{A} \pm \mathbf{B})^{(n + 1)}_{s_is_a}
    = \sum_{s_j,s_b}\left\{ - W(\omega=0)^{(n)}_{s_as_b,s_js_i} \mp W(\omega=0)^{(n)}_{s_as_j,s_bs_i} \right\}b^{(n)}_{s_is_a} \;.
\end{equation}
In the minus case, this is equivalent to the evaluation of the greater or lesser component of self-energy for a single imaginary time point. In the plus case, a similar algorithm can be used, but the resulting matrix needs to be antisymmetrized. We solve \eqref{davidson-matrix} in the basis of Slater functions and then transform to the subspace basis functions. For detailed working equations, we refer to appendix~\ref{app:working-eq}.

A key element in our approach is to use Pair-atomic density fitting (PADF)\cite{Krykunov2009, Merlot2013, Ihrig2015, Wirz2017, Forster2020, Forster2020b} to calculate the transformation from auxiliary basis to primary basis and back. in PADF, all the coefficients in the transformation matrix corresponding to auxiliary functions which are not centered on the same atoms as the primary basis functions are restricted to zero. While making the resulting basis transformation very efficient this also is an approximation which does not necessarily conserve important properties of the original matrices, like for example positive definiteness of the Coulomb potential.\cite{Wirz2017} These deficiencies can always be traced back to products of diffuse Slater functions which are difficult to expand in the auxiliary basis. To overcome these issues we introduce a projection technique to remove problematic linear dependencies from the primary basis which is described in appendix~\ref{app:lin-dependencies}.

\section{\label{sec:computational_details}Computational Details}\protect

All calculations have been performed with a locally modified development version of ADF2022.1\cite{Snijders2001,adf2022} The $GW$ implementation is the same as outlined in refs.~\citen{Forster2020b, Forster2021, Forster2021a}, except for the modification outlined in appendix~\ref{app:lin-dependencies}. 

For the hexameric unit of PSII, we used the structure of ref.~\citen{Sirohiwal2022} which has been optimized at the PBE level of theory taking into account environment effects using a QM/MM approach. Dimer structures have been optimized in this work using CAM-B3LYP-D3(BJ), a triple-$\zeta$ + polarization (TZP)\cite{vanLenthe2003} basis set and \emph{Good} numerical quality. The monomer structures used in section~\ref{app:starting-point} and sec.~\ref{app:basis-set} are taken from the structure by ref.~\citen{Kavanagh2020} based on the experimental structure at 1.9 {\AA} resolution by Umena et al.\cite{Umena2011} and where the positions of the Hydrogen atoms have been optimized using a semi-empirical model with all other coordinates frozen. All structures used in this work can be found in the supporting information.

We also benchmarked the basis set dependence of the $GW$-BSE calculations using the larger TZ3P and QZ6P basis sets\cite{Forster2021} for Chla monomers in section~\ref{app:basis-set}. All qs$GW$-BSE calculations reported in table~\ref{tab:basis-set} have been obtained with the \emph{veryGood} auxiliary basis. This allows us to reliably compare excitation energies obtained with different primary basis sets. TZ3P and QZ6P contain $f$-functions for second-row atoms and for such basis sets, the \emph{Good} auxiliary fit set is generally insufficient. For monomers, we calculate the lowest 6 eigenstates of \eqref{generalized}.

For chromophore dimers we calculated the lowest 6 eigenstates of \eqref{generalized}, using TZP (triple-$\zeta$ + polarization)\cite{vanLenthe2003} as primary basis set, \emph{Good} numerical quality and 16 imaginary time and frequency points each. In all calculations for monomers and dimers we terminate the sequence of subspace iterations if all eigenvalues are converged within $10^{-5}$ Hartree (0.27 meV).

In the $GW$-BSE calculations of the excited states of the hexamer, we used the TZP basis set, \emph{Basic} numerical quality, and 12 imaginary time and frequency points each. We restrict the basis in which we solve the BSE to the subspace spanned by all particle-hole pairs with transition energies below 1.5 Hartree. In agreement with earlier $GW$-BSE studies for such systems,\cite{Faber2013} we found this approximation to change the low-lying excitation energies by only around 10-20 meV compared to calculations including all particle-hole pairs.\bibnote{For instance, changing the cut-off for the inclusion of the particle-hole states from 1.5 to 2.0 Hartree changes each of the lowest three excitation energies of monomers by less than 10 meV.} This improves numerical stability of our algorithm and accelerates the convergence of the subspace iterations in the Davidson algorithm. We perform eight subspace iterations in the Davidson algorithm and calculate the 24 lowest eigenstates of \eqref{generalized}. This is sufficient to converge the low-lying excited states to within less than 5 meV. We also calculated the low-lying excited states of the same system using TD-DFT with the $\omega$B97-X kernel using the same numerical settings. However, in contrast to our $GW$-BSE calculations, we calculated the 12 lowest states and converged all eigenvalues to within $10^{-6}$ Hartree.

In all calculations we took into account scalar relativistic effects in the zeroth-order regular approximation.\cite{VanLenthe1993,VanLenthe1994, VanLenthe1996} The threshold $\epsilon_s$ described in appendix~\ref{app:lin-dependencies} has been set to $5 \times 10^{-3}$. Also, in all KS calculations we set the threshold below which we set eigenvalues of the inverse of the overlap matrix to zero during he canonical orthonormalization procedure to $5 \times 10^{-3}$.  If not stated otherwise, in all qs$GW$ calculations we first perform a PBE0 calculation with 40 \% exact exchange (PBEH40), which is a good preconditioner for qs$GW$ and leads to fast convergence.\cite{Belic2022} 
Aside from numerical inaccuracies, the final results are independent of this choice which we have verified in ref.~\citen{Forster2021a} and which we will verify also for the case of Chla in the next section. For qs$GW$, we terminate the calculations when the Frobenius norm of the difference between the density matrices of two subsequent iterations falls below $5 \times 10^{-9}$.\cite{Forster2021a} We also performed ev$GW$-BSE calculations based on the LDA and PBEH40 functionals (ev$GW@$LDA, ev$GW@$PBEH40). We terminate the ev$GW$ calculations if the HOMO QP energy difference between two subsequent iterations falls below 3 meV.

To compare our method to the RSH TD-DFT approach, we also performed calculations using the CAMY-B3LYP and $\omega$B97-X kernel using the TZP basis set and \emph{Good} numerical quality. We also calculated the electrochromatic shifts due to the presence of the protein environment using the conductor like screening model (COSMO)\citep{Klamt1993, Klamt1995, Klamt1996} as implemented in ADF.\citep{Pye1999} Following ref.~\citen{Suomivuori2016}, we set the dielectric constant of the environment to a value of $4.0$ in these calculations which should approximately account for solvent and protein environment.  

\section{\label{sec:results}Results}\protect

\begin{figure}[hbt!]
    \centering
    \includegraphics[width=0.9\textwidth]{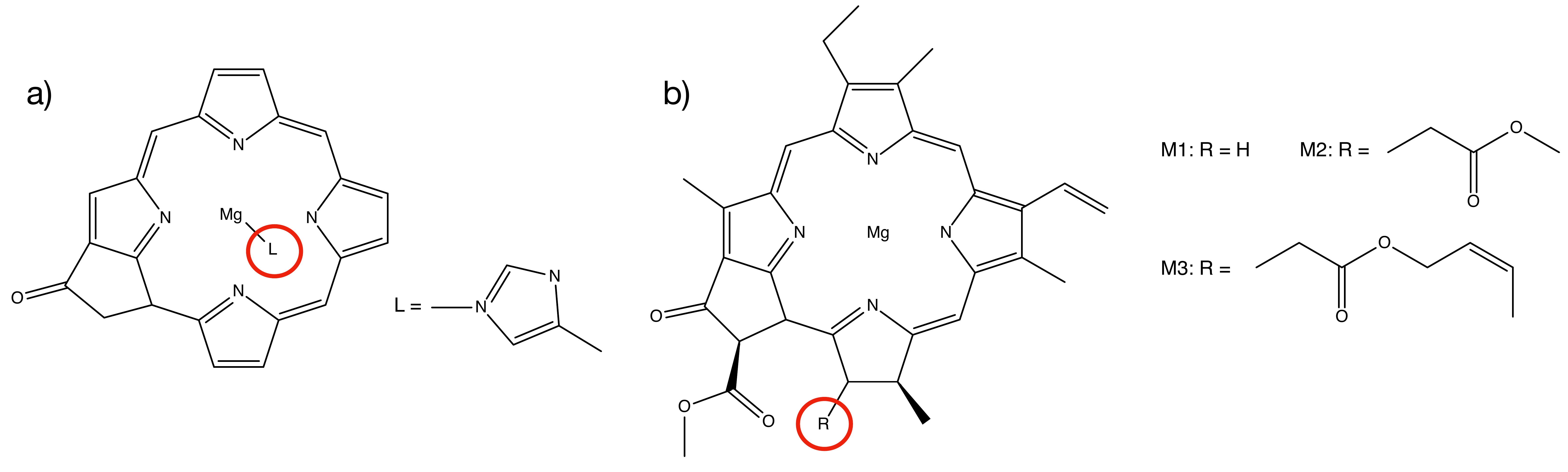}
    \caption{Different models of Chla used in this work: a) Model used by Suomivuori et al.\cite{Suomivuori2016} with ligating Histidine residue. b) Models without Histidine residue but containing all ligands at the chlorin core and different models for the phytyl chain (M1, M2, and M3, respectively).}
    \label{fig:structures}
\end{figure}

\subsection{\label{app:starting-point}Starting-point dependence}
As discussed in the introduction, its starting point independence is a major advantage of qs$GW$ over ev$GW$. To verify the starting point independence of our implementation, we report here vertical excitation energies (VEE) for qs$GW$ and ev$GW$ for the M2 model in figure~\ref{fig:structures}b) with 82 atoms in total for the LDA, PBE, PBEH40, and HF  starting points. We thereby use a tighter convergence criterion of 1 meV for the HOMO QP energy for ev$GW$ than the default value. The results for the $Q_y$ excitation are shown in table~\ref{tab:starting-point}. The qs$GW$ calculations converge to the same HOMO-LUMO gap within an accuracy of 10 meV within less than 10 iterations. This also results in $Q_y$ excitation energies which are converged within 10 meV. The remaining differences are due to numerical noise in the imaginary frequency and time grids used in the $GW$ calculations which then translates into uncertainties in the analytical continuation of the self-energy to the complex plane.\cite{Forster2021, Wilhelm2021} The differences in the HOMO-LUMO gaps of the ev$GW$ calculations are much larger and differ by almost 300 meV between ev$GW@$LDA and ev$GW@$HF, which results in $Q_y$ excitations energies differing by about 80 meV. This is the most extreme case, for starting points other than HF there are only very small differences between the different ev$GW$ results. This has already been observed in ref.~\citen{Hashemi2021}. Since the computational overhead of a qs$GW$ calculation is negligible compared to ev$GW$ (5.79 vs. 5.67 core hours per iteration) and the number of iterations needed for convergence is essentially the same, there is little advantage to be gained by using ev$GW$ instead of the more robust qs$GW$ approach.

\begin{table*}[hbt!]
    \centering
    \begin{tabular}{lcccccccc}
    \toprule
    & \multicolumn{4}{c}{qs$GW$}
    & \multicolumn{4}{c}{ev$GW$} \\
    & gap & $Q_y$ [eV] & $n_I$ & $t$ [h]
    & gap & $Q_y$ [eV] & $n_I$ & $t$ [h] \\
    \midrule
         LDA    &  4.499 & 1.752 & 9  & 5.79 & 4.405 & 1.764 & 9 & 5.67 \\
         PBE    &  4.501 & 1.745 & 10 &  -   & 4.417 & 1.837 & 9 & - \\
         PBEH40 &  4.493 & 1.760 &  8 &  -   & 4.476 & 1.772 & 7 & - \\ 
         HF     &  4.496 & 1.753 &  9 &  -   & 4.671 & 1.766 & 9 & - \\
    \bottomrule
    \end{tabular}
    \caption{HOMO-LUMO gap, Value of the $Q_y$ excitation for different starting points, number of iterations until convergence and time per $GW$ iteration, measured in core hours, for qs$GW$ and ev$GW$. Calculations were performed on a 2.2 GHz intel Xeon (E5-2650~v4) node (broadwell architecture) with 24 cores and 128 GB RAM.}
    \label{tab:starting-point}
\end{table*}

\subsection{\label{app:basis-set}Basis Set Errors} 

\begin{table*}[hbt!]
    \centering
    \begin{tabular}{lcccccccccccc}
    \toprule
    & \multicolumn{6}{c}{ev$GW@$LDA-BSE} &
    \multicolumn{6}{c}{qs$GW$-BSE} \\
    & \multicolumn{3}{c}{M1} &
    \multicolumn{3}{c}{M2} &
    \multicolumn{3}{c}{M1} &
    \multicolumn{3}{c}{M2} \\
    \cline{2-4} \cline{5-7} \cline{8-10} \cline{11-13}
    & $Q_y$ & $Q_x$ & $B$ & $Q_y$ & $Q_x$ & $B$ & $Q_y$ & $Q_x$ & $B$ & $Q_y$ & $Q_x$ & $B$ \\ 
    \midrule
    TZP           & 1.74 & 1.93 & 2.68 & 1.76 & 1.94 & 2.71 & 1.72 & 1.98 & 2.84 & 1.74 & 2.00 & 2.86 \\
    TZ3P          & 1.77 & 1.96 & 2.72 & 1.79 & 1.98 & 2.76 & 1.72 & 1.98 & 2.84 & 1.73 & 1.97 & 2.84 \\
    QZ6P          & 1.71 & 1.94 & 2.64 & 1.74 & 1.92 & 2.68 & 1.71 & 1.96 & 2.80 & 1.71 & 1.96 & 2.84 \\
    $\Delta_{TQ}$ & 0.03 &-0.01 & 0.04 & 0.02 & 0.02 & 0.03 & 0.01 & 0.02 & 0.04 & 0.03 & 0.04 & 0.02 \\
    \bottomrule
    \end{tabular}
    \caption{VEEs for M1 and M2 with different basis sets for qs$GW$-BSE and ev$GW@$LDA-BSE. The values in the last row denote the differences in VEEs calculated with the TZP\cite{vanLenthe2003} and QZ6P\cite{Forster2021} basis sets. All values are in eV.}
    \label{tab:basis-set}
\end{table*}

Next, we investigate the dependence of the $Q_y$ excitation energy on the basis set size. For $GW$ calculations it is well known that individual QP energies converge slowly with respect of the size of the single-particle basis. In practice extrapolation techniques are needed to obtain converged results.\cite{VanSetten2015, Stuke2020, Bruneval2020} For orbital energy differences which are entering the BSE, the situation is much better since the basis set error for the QP energies usually have the same sign.\cite{Stuke2020} In table~\ref{tab:basis-set} we compare the lowest excitation energies calculated with different basis sets for the two different Chla models M1 and M2 shown in figure~\ref{fig:structures}b). For ev$GW$ and qs$GW$ the QZ6P VEEs are only slightly lower than the TZP ones, indicating that they are almost converged also with the smaller basis set. These errors are certainly smaller then other possible sources of error in our calculations like shortcomings of $GW$-BSE or uncertainties in structural parameters. Therefore, to a very good approximation, we can ignore the basis set incompleteness error in all of the following TZP calculations. 

\subsection{Comparison to Experiment and different \emph{ab-initio} Calculations} 

\subsubsection{Monomers}

Next, we assess the accuracy of qs$GW$-BSE by comparison to experimental gas-phase data for Chla by Gruber et al.\cite{Gruber2019} In table~\ref{tab:cla} we directly compare VEEs calculated with different computational methods to the experimental VEE which has recently been extracted from the experimental spectrum by Sirohiwal et al.\cite{Sirohiwal2020}. The domain based local pair-natural orbital\cite{Riplinger2013, Riplinger2013a} (DLPNO)-STEOM-CCSD\cite{Dutta2016, Dutta2017, Dutta2019} results are taken from ref.~\citen{Sirohiwal2020}, while the ev$GW@$LDA-BSE/6-311++G(2d,2p) results calculated using MOLGW\cite{Bruneval2016a} are by Hashemi and Leppert.\cite{Hashemi2021} Two different, gas-phase optimized structures have been used: One has been optimized at the CAM-B3LYP-D3(BJ)/def2-TZVP level of theory by Sirohival et al.\cite{Sirohiwal2020}, while the other has been optimized by Hashemi and Leppert using B3LYP/def2-TZVP. 

\begin{table*}[hbt!]
    \centering
\begin{tabular}{lcccc}
\toprule 
         & $Q_y$ & $Q_x$ & $B$ & $\Delta_{Q_y - Q_x}$ \\ 
\midrule 
    exp. (VEE)      & 1.99 & 2.30 & 3.12 & 0.31 \\ 
    exp. (band max) & 1.94 & 2.23 & 3.08 & 0.29 \\ 
    \multicolumn{5}{c}{CAM-B3LYP-D3(BJ)/def2-TZVP optimized structure} \\
\midrule 
    DLPNO-STEOM-CCSD  & 1.75 & 2.24 & 3.17 & 0.49 \\ 
    qs$GW$            & 1.97 & 2.29 & 3.15 & 0.32 \\
    ev$GW@$PBEH40     & 1.98 & 2.29 & 3.15 & 0.31 \\ 
    ev$GW@$LDA        & 1.94 & 2.20 & 3.01 & 0.26 \\ 
    CAMY-B3LYP        & 1.94 & 2.23 & 3.08 & 0.29 \\ 
    $\omega$B97-X     & 2.10 & 2.71 & 3.57 & 0.61 \\ 
    \multicolumn{5}{c}{B3LYP/def2-TZVP optimized structure} \\
\midrule 
    ev$GW@$LDA-BSE (ADF/TZP)               & 1.85 & 2.09 & 2.91 & 0.24 \\    
    ev$GW@$LDA-BSE (MOLGW/6-311++G(2d,2p)) & 1.85 & 2.13 & 2.91 & 0.28 \\
\bottomrule 
\end{tabular}
    \caption{VEEs for Chla calculated with different quantum chemical methods for two different gas-phase optimized structures and experimental reference data. All values are in eV.}
    \label{tab:cla}
\end{table*}

We performed ev$GW@$LDA-BSE calculations for both structures. Our results for the CAM-B3LYP-D3(BJ) optimized structure are consistently around 0.1 eV lower than the ones for the B3LYP optimized structure. This illustrates the large influence of small changes in structural parameters on the final excitation energies. However, CAM-B3LYP has been shown to describe the structural features of Cholorpyll monomers very well.\cite{Zamzam2019, Sirohiwal2020} For the B3LYP optimized structure, we can compare our herein calculated VEEs to the ones from Hashemi and Leppert calculated on the same level of theory. Except for the $Q_x$ excitation energies which are slightly different (40 meV), we find perfect agreement between both implementations. 

All ev$GW$ results agree very well with qs$GW$ also for Chla. All $GW$-BSE results for the CAM-B3LYP-D3(BJ) optimized structure are in excellent agreement with the experimental values. For instance, the qs$GW$-BSE VEEs agree all with the experimental VEEs within 30 meV. On the other hand, DLPNO-STEOM-CCSD not only severly underestimates the $Q_y$ excitation energy, but it also overestimates the gap between both $Q$-bands, $\Delta_{Q_y - Q_x}$, considerably. 
Considering this difference, we note that STEOM-CCSD is not necessarily a reliable reference for qs$GW$. In STEOM-CCSD, a much larger number of diagrams is considered in the single- and two-particle Green's functions compared to $GW$.\cite{Lange2018} QP approximations to $GW$ approximate the effect of these diagrams instead by neglecting the vertex.\cite{Kotani2007} The diagrams contained in $GW$ are not a subset of the ones contained in EOM-CCSD but only of the ones contained in EOM-CCSDT.\cite{Lange2018} Accounting for excitations to triples (at least to some extent) is known to be of high importance for the reliable description of charged\cite{Ranasinghe2019} and neutral excitations.\cite{Loos2018a, Loos2020e,Monino2021} Consequently, STEOM-CCSD shows mean signed errors compared to EOM-CCSDT calculations of around 0.1 eV for a set of medium organic molecules, but errors can be as large as 0.5 eV in some cases.\cite{Loos2020e} Moreover, apart from the neglect to triple excitations, the DLPNO approximation can also introduce some artifacts. The pairs which are treated on the CC level are selected based on an MP2 calculation\cite{Riplinger2013a} which is not always reliable for systems with strongly screened electron-electron interactions.\cite{Mattuck1992, Nguyen2020}

Lastly, TD-DFT with the RSH kernels CAMY-B3LYP and $\omega$B97-X which are typically used in computational studies of the PSII RC\cite{Frankcombe2015, Kavanagh2020, Sen2021, Sirohiwal2022} give very different results. CAMY-B3LYP is actually in excellent agreement with experiment and the $GW$-BSE calculations, while $\omega$B97-X gives much too high excitation energies and also massively overestimates the $\Delta_{Q_y - Q_x}$.

\subsubsection{Dimers}

\begin{table}[hbt!]
    \centering
    \begin{tabular}{lcccccc}
    \toprule
    kernel & $\Omega_{1}$ & $\Omega_{2}$ & $\Omega_{3}$
    & $\Omega_{4}$ & $\Omega_{5}$ & $\Omega_{6}$ \\
\midrule
    exp. (VEE)\cite{Milne2016}      & \multicolumn{6}{c}{1.95 (estimated)}  \\ 
    exp. (band max)\cite{Milne2016} & \multicolumn{6}{c}{1.90} \\ 
    \multicolumn{6}{c}{B3LYP-D3(BJ)/def2-SVP optimized structure$^a$\cite{Suomivuori2016}} \\
\midrule 
        ev$GW@$LDA        & 1.87 & 1.88 & 1.90 & 1.90 & 2.72 & 2.75 \\
        ev$GW@$PBEH40     & 1.92 & 1.95 & 2.09 & 2.11 & 2.84 & 2.93 \\
        qs$GW$            & 1.89 & 1.92 & 2.07 & 2.10 & 2.83 & 2.92 \\
        CAMY-B3LYP        & 2.12 & 2.15 & 2.29 & 2.32 & 2.63 & 2.76 \\
        RVS-LT-SOS-ADC(2)$^b$ & 2.04 & 2.06 \\
    \multicolumn{6}{c}{CAM-B3LYP-D3(BJ)/TZP optimized structure$^{c}$} \\
\midrule 
        ev$GW@$LDA        & 1.98 & 1.99 & 2.16 & 2.22 & 2.51 & 2.64 \\
        ev$GW@$PBEH40     & 1.97 & 2.02 & 2.24 & 2.27 & 2.58 & 2.67 \\
        qs$GW$            & 1.94 & 1.98 & 2.25 & 2.28 & 2.56 & 2.68 \\
        CAMY-B3LYP        & 2.12 & 2.16 & 2.38 & 2.43 & 2.51 & 2.61 \\
        $\omega$B97-X     & 2.05 & 2.10 & 2.63 & 2.68 & 3.10 & 3.27 \\
        \bottomrule
    \end{tabular}
    \caption{The lowest six excitation energies for two different models of the Chla dimer. All values are in eV.$^{a,b,c}$}
    \label{tab:cla_dim}
    \footnotesize{$^a$The B3LYP-D3(BJ)/def2-SVP structure has been taken from Suomivuori et al.\cite{Suomivuori2016}.} \\
    \footnotesize{$^b$Results taken from Suomivuori et al.\cite{Suomivuori2016}.} \\
    \footnotesize{$^c$The structure of the M3 dimer has been optimized in this work at CAM-B3LYP-D3(BJ)/TZP.}
\end{table}

In table~\ref{tab:cla_dim}, we show the low-lying excitations of $GW$-BSE calculations for different models of P\textsubscript{D1}-P\textsubscript{D2}. The first dimer structure has been optimized in the gas phase by Suomivuori \emph{et al.} at the B3LYP-D3/def2-SVP level of theory and consists of two Chla monomers whose structure is shown in figure~\ref{fig:structures}a. This structure lacks most substituents of the Chlorin core present in Chla (see figure~\ref{fig:structures}b which, in principle, complicates comparison of excitation energies to experimental results. However, these calculations give some indication on the performance of $GW$-BSE in comparison to the RVS-LT-SOS-ADC(2) VEEs by Suomivuori \emph{et al.} Comparison of experimental band maximum and VEE for a single Chla measured in ref.~\citen{Gruber2019} suggests that the VEE of the chlorophyll dimer might be around 1.95 eV (50 meV higher than the band maximum).

\begin{table*}[hbt!]
    \centering
    \begin{tabular}{lcccccccc}
    \toprule
    & \multicolumn{4}{c}{ev$GW@$LDA} & \multicolumn{4}{c}{ev$GW@$PBEH40} \\
    & VEE & character & weight & $f$             & VEE  & character  & weight & $f$ \\
    \midrule
     $\Omega_1$ & 1.87 & 238 $\rightarrow$ 240 & 0.49 & 0.08 & 1.92 & 238 $\rightarrow$ 240 & 0.28 & 0.30 \\
                      &      &                       &      &      &      & 237 $\rightarrow$ 239 & 0.26 &      \\ 
     $\Omega_2$ & 1.88 & 237 $\rightarrow$ 240 & 0.22 & 0.14 & 1.95 & 238 $\rightarrow$ 241 & 0.41 & 0.03 \\
                      &      & 237 $\rightarrow$ 239 & 0.17 &      &      & 237 $\rightarrow$ 239 & 0.34 &      \\ 
     $\Omega_3$ & 1.90 & 236 $\rightarrow$ 239 & 0.38 & 0.13 & 2.09 & 235 $\rightarrow$ 239 & 0.53 & 0.04 \\
     $\Omega_4$ & 1.90 & 237 $\rightarrow$ 240 & 0.37 & 0.00 & 2.11 & 236 $\rightarrow$ 240 & 0.49 & 0.03 \\
                      &      & 235 $\rightarrow$ 239 & 0.31 &      &      &                       &      &      \\
     $\Omega_5$ & 2.72 & 238 $\rightarrow$ 239 & 0.51 & 0.37 & 2.84 & 238 $\rightarrow$ 239 & 0.56 & 0.24 \\
     $\Omega_6$ & 2.75 & 237 $\rightarrow$ 239 & 0.27 & 0.14 & 2.93 & 237 $\rightarrow$ 240 & 0.31 & 0.20 \\
                      &      & 237 $\rightarrow$ 242 & 0.24 &      &      &                       &      &      \\
    \bottomrule
    \end{tabular}
    \caption{Characterization and comparison of the low-lying excited states of Chla dimer (structure by Suomivuori et al.\cite{Suomivuori2016}) calculated with ev$GW@$LDA-BSE and ev$GW@$PBEH40-BSE.$^a$}
    \label{tab:cla_dim_character}
    \footnotesize{$^a$Shown are the excitation energies $\Omega_{S}$ (in eV), the dominant coefficients of the corresponding eigenvector and the associated particle-hole transitions, as well as the oscillator strengths $f$.}
\end{table*}

\begin{figure}[hbt!]
    \centering
    \includegraphics[width=0.9\textwidth]{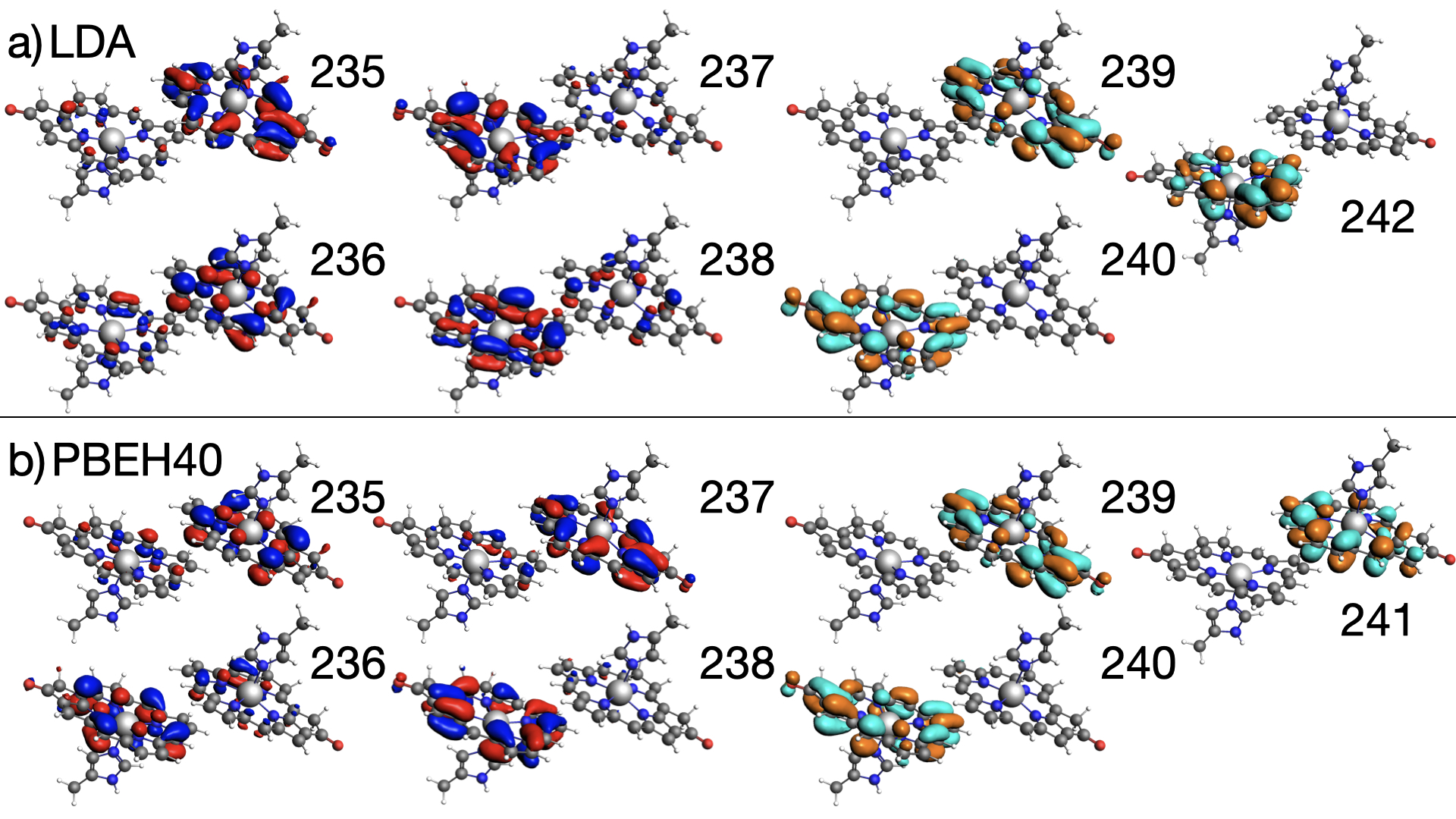}
    \caption{Selected valence single-particle KS orbitals for the Chla dimer (structure by Suomivuori et al.\cite{Suomivuori2016}) calculated using LDA and PBEH40.}
    \label{fig:orbitals_KS_dim108}
\end{figure}

As for the monomer, the $GW$-BSE results are in excellent agreement with these values while the RVS-LT-SOS-ADC(2) VEEs are much too high. In contrast to the case of the Chla monomer, CAMY-B3LYP  overestimates the VEEs by far. The VEEs $\Omega_{3}$ and $\Omega_4$ of the BSE calculation based on ev$GW@$LDA are almost 0.2 eV lower than the ones based on ev$GW@$PBEH40, and in the former calculation, the four lowest excited states are almost degenerate. The character of these excitations are compared in more detail in table~\ref{tab:cla_dim_character} with the corresponding KS single-particle orbitals shown in figure~\ref{fig:orbitals_KS_dim108}. Comparison of the most important contributions to the eigenvector $\ket{\mathbf{X},\mathbf{Y}}^T_1$ already shows that ev$GW@$LDA-BSE predicts the lowest excitation to be localized on the P\textsubscript{D1} fragment, while in the ev$GW@$PBEH40-BSE calculation it is delocalised over both monomers with almost equal weights. Using ev$GW@$LDA-BSE, the second excited state has a large contribution of a particle-hole transition located on P\textsubscript{D1}, while it is localized on P\textsubscript{D2} using ev$GW@$PBEH40-BSE. Also, the oscillator strengths in table~\ref{tab:cla_dim_character} show that the different excitations differ substantially in their brightness. Together with the large difference in some of the VEEs, this shows that different KS starting points can lead to different excitations, even when the eigenvalues are updated self-consistently. 

In table~\ref{tab:cla_dim}, we also show results for a more realistic model of the Chla dimer. Our model consists of two M3 monomers which includes the first four segments of the phytyl chain in stacked conformation. In table S1 of the supporting information, we show that the final excitation energies are however very insensitive to the particular structural model. 

The band maximum of ref.~\citen{Milne2016} which we used as reference has been measured for a charge tagged dimer. However, as shown in ref.~\citen{Milne2015} for Chla monomers, the final excitation energies are insensitive to the type of charge tag and omitting the charge tag entirely only results in a lowering of the excitation energies of around 30-40 meV.

The excitations have been calculated for a geometry optimized at the CAM-B3LYP/TZP level of theory. Excitation energies for geometries optimized with different methods can be found in table S2 of the supporting information. In accordance with ref.~\citen{Sirohiwal2020} and our results shown in table~\ref{tab:cla} we found the VEEs to be very sensitive to the choice of the functional chosen for geometry optimization. For instance, using PBE-D4/TZP lowers the lowest 2 excitation energies by around 0.1 eV with respect to the CAM-B3LYP-D3(BJ) optimized structure. The data shown in table S3 in supporting information furthermore demonstrates that VEEs for crystal structures considerably underestimate the experimental values.

The lowest qs$GW$-BSE excitation energy of 1.94 meV is again in excellent agreement with the VEE of 1.95 eV estimated from the band maximum. As explicitly shown in the supporting information and as for the monomers in table~\ref{tab:basis-set}, the excitation energies are again rather insensitive to the basis set.
Also notice that the remaining small basis set errors will largely cancel with the small error from omitting the charge tag. Again, the lowest two ev$GW$-BSE VEEs are in excellent agreement with the qs$GW$-BSE one and each other, while there are larger differences in higher-lying VEEs. As for the monomer, CAMY-B3LYP massively overestimates the VEEs compared to experiment.

\subsection{Six-chromophore model of the PSII RC}

\begin{table}[h!]
    \centering
    \begin{tabular}{lcclc} 
    \toprule
     & VEE & $f$ & Character & weight \\
    \midrule 
     \multirow{2}{*}{$\Omega_1$} & \multirow{2}{*}{1.89} & \multirow{2}{*}{0.22} & P\textsubscript{D2}$^*$  & 0.39 \\
     & & & Chl\textsubscript{D2}$^*$ & 0.22 \\
     \midrule 
     \multirow{3}{*}{$\Omega_3$} & \multirow{3}{*}{1.90} & \multirow{3}{*}{0.77} & 
     P\textsubscript{D2}$^*$ & 0.24 \\
     & & & P\textsubscript{D1}$^*$ & 0.14 \\
     & & & Pheo\textsubscript{D2}$^*$ & 0.09 \\ 
     & & & P\textsubscript{D1}$^+$ - P\textsubscript{D2}$^-$ & 0.09 \\
     \midrule 
     \multirow{3}{*}{$\Omega_3$} & \multirow{3}{*}{1.91} & \multirow{3}{*}{0.04} & 
     Chl\textsubscript{D1}$^*$ & 0.30 \\
     & & & P\textsubscript{D1}$^*$ & 0.24 \\
     & & & Chl\textsubscript{D1}$^+$ - Pheo\textsubscript{D1}$^-$   & 0.08 \\   
     \midrule
     \multirow{4}{*}{$\Omega_4$} & \multirow{4}{*}{1.92} & \multirow{4}{*}{0.22} & 
     Pheo\textsubscript{D2}$^*$      & 0.39 \\   
     & & & Chl\textsubscript{D2}$^*$  & 0.16 \\
     & & & Pheo\textsubscript{D2}$^*$ & 0.12 \\
     & & & Chl\textsubscript{D1}$^*$  & 0.09 \\
     \midrule 
     \multirow{4}{*}{$\Omega_5$} & \multirow{4}{*}{1.94} & \multirow{4}{*}{0.01} & 
     Chl\textsubscript{D1}$^*$        & 0.23 \\   
     & & & Chl\textsubscript{D2}$^*$  & 0.18 \\
     & & & P\textsubscript{D1}$^*$   & 0.16 \\
     & & & P\textsubscript{D2}$^*$   & 0.15 \\
     \midrule 
     \multirow{2}{*}{$\Omega_6$} & \multirow{2}{*}{1.97} & \multirow{2}{*}{0.20} & 
     Pheo\textsubscript{D1}$^*$ & 0.54 \\
     &&&Pheo\textsubscript{D1}$^-$ - Chl\textsubscript{D1}$^+$ & 0.21 \\
          \midrule 
     \multirow{2}{*}{$\Omega_{13}$} & \multirow{2}{*}{2.71} & \multirow{2}{*}{0.00} & 
     P\textsubscript{D2}$^+$ - Chl\textsubscript{D2}$^-$ & 0.81 \\
     &&&P\textsubscript{D1}$^+$ - Chl\textsubscript{D2}$^-$ & 0.13 \\
          \midrule 
     \multirow{2}{*}{$\Omega_{14}$} & \multirow{2}{*}{2.73} & \multirow{2}{*}{0.00} & 
     P\textsubscript{D1}$^+$ - Chl\textsubscript{D1}$^-$ & 0.70 \\
     &&&P\textsubscript{D1}$^+$ - Pheo\textsubscript{D1}$^-$ & 0.20 \\
    \bottomrule
    \end{tabular}
    \caption{The lowest qs$GW$@-BSE/TZP excited states of the hexameric chromophore complex in the RC of PSII.${^a}$.}
    \footnotesize{$^a$Shown are the excitation energies $\Omega_{S}$ (in eV), the dominant coefficients of the corresponding eigenvector and the associated particle-hole transitions, as well as the oscillator strengths $f$.}
    \label{tab::hexamer}
\end{table}

The most complete model of the PSII RC we consider in this work comprises all six chromophores shown in figure~\ref{fig::chromophores} with 476 atoms in total. Time-resolved spectroscopic experiments\cite{Groot2005, Romero2012,Yoneda2022} show that the primary electron transfer in the RC occurs from an exciton localized on Chl\textsubscript{D1} to Pheo\textsubscript{D1}, followed by a transfer of the hole to P\textsubscript{D1}. This would point to the mixing in of low-lying CT states with pronounced Chl\textsubscript{D1}\textsuperscript{+} -Pheo\textsubscript{D1}\textsuperscript{-} and P\textsubscript{D1}\textsuperscript{+} -Pheo\textsubscript{D1}\textsuperscript{-} character in calculations of excitation energies. In previous TD-DFT calculations using RSH kernels for similar multi-chromophoric models, no low-lying CT state which could be related to this charge separation pathway have been observed.\cite{Frankcombe2015,Sirohiwal2022} In recent computational studies, both Sirohiwal et al.\cite{Sirohiwal2020a, Sirohiwal2022} and Tamura et al.\cite{Tamura2021} demonstrated that the protein environment is crucial for observing the Chl\textsubscript{D1}\textsuperscript{+} -Pheo\textsubscript{D1}\textsuperscript{-} CT state at low energies.

\begin{table}[hbt!]
    \centering
    \begin{tabular}{lcccccc}
    \toprule 
    & \multicolumn{2}{c}{qs$GW$-BSE} & \multicolumn{2}{c}{qs$GW$@PBEH40-BSE}
    & \multicolumn{2}{c}{TD-DFT@$\omega$B97-X} \\ 
    & VEE & $f$ & VEE & $f$ & VEE & $f$ \\ 
    \midrule 
       {$\Omega_1$}  & 1.89 & 0.22 & 1.94 & 0.81 & 1.92 & 0.33 \\
       {$\Omega_2$}  & 1.90 & 0.77 & 1.94 & 0.32 & 1.93 & 0.64 \\
       {$\Omega_3$}  & 1.91 & 0.04 & 1.96 & 0.05 & 1.94 & 0.14 \\
       {$\Omega_4$}  & 1.92 & 0.22 & 1.97 & 0.24 & 1.96 & 0.18 \\
       {$\Omega_5$}  & 1.94 & 0.01 & 1.99 & 0.15 & 1.97 & 0.09 \\
       {$\Omega_6$}  & 1.97 & 0.20 & 2.00 & 0.11 & 1.98 & 0.07 \\
        \bottomrule
    \end{tabular}
    \caption{The VEEs and oscillator strengths of the six lowest excited states of the hexameric complex at different levels of theory. All values are in eV.}
    \label{tab::compare_hexamer}
\end{table}

The low-lying excitations of the hexameric complex at the qs$GW$-BSE/TZP level of theory are characterized in table~\ref{tab::hexamer}. In the supporting information we characterize these excitations in more detail by visualzing the involved single-particle qs$GW$ orbitals. We also present results of our own TD-DFT calculations using the $\omega$B97-X kernel as well as for ev$GW$@PBEH40-BSE/TZP. The excitation energies and the oscillator strengths of the six lowest excited states using these different methods are compared in table~\ref{tab::compare_hexamer}. 

In agreement with past\cite{Frankcombe2015, Sirohiwal2022} and our own TD-DFT calculations using the $\omega$B97-X kernel, only states with local character can be found among the six lowest excitations of the hexamer using both, qs$GW$-BSE and ev$GW$@PBEH40-BSE. As shown in table~\ref{tab::compare_hexamer}, also the VEEs using the different methods agree within 50 meV. In all methods, the low-lying states are linear combinations of excitonic states involving the frontier orbitals on each chromophore. 

At the qs$GW$-BSE level, the two lowest states with pronounced CT character can be found at 2.7 eV and these cannot directly be linked to charge separation pathways in PSII which have been observed experimentally.\cite{Groot2005, Romero2012, Yoneda2022} Only the third excited state at the qs$GW$-BSE level of theory at 1.91 eV contains a contribution from a Chl\textsubscript{D1}\textsuperscript{+} -Pheo\textsubscript{D1}\textsuperscript{-} particle-hole transition with a small weight, which is entirely absent in out TD-DFT and ev$GW$-BSE calculations. Future studies at the $GW$-BSE level with inclusion of the environment electrostatics are needed to rationalize how the Chl\textsubscript{D1}\textsuperscript{+} -Pheo\textsubscript{D1}\textsuperscript{-} CT state is influenced by the protein environment at the qs$GW$-BSE level. 

\subsection{\label{app:timings}Timings} 

\begin{table*}[hbt!]
    \centering
    \begin{tabular}{lcrcccccc}
    \toprule 
    & & & &
    \multicolumn{2}{c}{Iterations} & \multicolumn{3}{c}{CPU time} \\
   Method & Basis & $N_{\text{bas}}$  & $N_{\Omega}$ & qs${GW}$ & BSE &
     $GW$ & BSE  & total \\
    \midrule
      \multirow{2}{*}{qs$GW$-BSE} 
       & TZ3P &11116 & 12 & 6  &10 & 3401 & 3447 & 7283 \\
       & TZP  & 6256 & 24 & 6  & 8 & 1074 & 1729 & 2924 \\
      ev$GW$-BSE 
      & TZP & 6256   & 24 & 5  & 8 & 826 & 1969 & 2917 \\
      $\omega$B97-X 
      & TZP & 6256 & 12 & --  & 21 & -- & 2675 & 2846 \\
    \bottomrule
    \end{tabular}
    \caption{CPU times (in core hours) to calculate the $N_{\Omega}$ lowest roots of the full hexamer with 476 atoms and 1872 correlated electrons with different basis sets and methods. 39884 auxiliary basis functions have been used in all calculations. All calculations have been performed on an 2.6 GHz AMD Rome 7H12 node with 64 cores and 16 GB RAM per node.}
    \label{tab:cpu-hours}
\end{table*}
Finally, we briefly comment on the computational effort for different basis sets and methods to calculate the lowest $N_{\Omega}$ roots of the full hexamer with 476 atoms and 1872 correlated electrons. The computational timings in core hours are given in table~\ref{tab:cpu-hours}. The calculation for the hexamer can be performed in less than 3000 core hours, i.e. in less than two days on a node with 64 cores. The qs$GW$ part of the calculation is slightly less expensive as the BSE part. Notice, that the BSE part of the calculation is roughly es involved as the TD-DFT calculation with the $W$B97-X kernel if the timings are normalized by the number of states and number of subspace iterations in the Davidson algorithm.

Notice, that low-order scaling implementations like ours which rely on sparsity in the primary basis usually do not scale well with the size of the basis set, as can be seen by comparing the timings of the qs$GW$-BSE calculations with different basis sets. We also performed a qs$GW$ calculation for the full hexamer with more than 11000 basis functions using the TZ3P basis set. Here, a single qs$GW$ iteration already takes around 540 core hours, which is more than three times more than one iteration using the TZP basis set. While in this work the TZP basis set was already sufficient to obtain converged results, typically lager basis sets will be required. Finite basis set correction techniques for many-body perturbation theory might be a promising solution to circumvent this problem.\cite{Bruneval2016, Riemelmoser2020, Loos2020a, Bruneval2020}

For larger calculations, the bottleneck of the computation is the number of auxiliary fit functions $N_{\text{fit}}$ (almost 40,000 for the hexamer). When large basis sets are used large auxiliary fit sets are necessary to guarantee numerical stability in the PADF approach and also in related techniques which rely on sparse transformation between matrices in primary and auxiliary basis.\cite{Wilhelm2021, Duchemin2021a} For each imaginary time and frequency point, a matrix of size $N_{\text{fit}} \times N_{\text{fit}} \approx 14 $GB needs to be stored. This amounts to almost 500 GB for the hexamer and if we were to double the system size, 2 TB of distributed memory would be needed. In our current implementation, we store these matrices on disk and transferring them to the CPU and back becomes very time-consuming. 

\section{\label{sec:conclusion}Conclusions}\protect
So far, applications of the $GW$-BSE method have been limited to rather small molecules.\cite{Hashemi2021, Liu2020, Gui2018} We presented here a new implementation of the method which enables its routine application to much larger systems. As opposed to a recently developed simplified $GW$-BSE scheme,\cite{Cho2022} our implementation does not introduce any empirical approximations to the matrix elements of the BSE Hamiltonian. Our implementation allowed us to calculate the 12 lowest excited states of the complete complex of six chromophores in the PSII RC with almost 2000 correlated electrons on the qs$GW$-BSE/TZP level. The calculation with around 6000 primary basis functions could be performed in a little more than one day on a single compute node. 

Since the single-particle states are optimized self-consistently, making the results independent of a mean-field reference calculation, qs$GW$-BSE is a theoretically more rigorous approach than ev$GW$-BSE. qs$GW$-BSE calculations for optimized geometries are in excellent agreement with experimental VEEs in the gas phase for Chla monomers and dimers. We have shown here explicitly for Chla dimers that ev$GW$-BSE might lead to different excitations for different starting points. This is in contrast to the generally good agreement for different starting points for monomers\cite{Hashemi2021} and can be seen as a major shortcoming of ev$GW$-BSE. We therefore conclude, that self-consistency in the single-particle states is decisive for a reliable description of the low-lying excitonic states of large chromophoric complexes. 

In agreement with previous results and our own calculations on the TD-DFT/RSH level for the full hexameric complex\cite{Frankcombe2015} also ev$GW$-BSE and qs$GW$-BSE only predict states with predominantly local character in the absence of the protein environment. 
These states can however not be linked to experimentally observed CT processes.\cite{Groot2005, Romero2012, Yoneda2022}
Recent computational studies have established that the environment electrostatics are responsible for this type of CT.\cite{Sirohiwal2020a, Sirohiwal2022, Tamura2021} Along the lines of previous $GW$-BSE implementations,\cite{Duchemin2018, Tirimbo2020, Tirimbo2020a} future research therefore needs to focus on ways to explicitly account for the environment electrostatics in large-scale $GW$-BSE calculations.

\appendix

\section{\label{sec::shifts}Electrochromatic shifts}

 \begin{table*}[hbt!]
     \centering
     \begin{tabular}{lccccccc}
     \toprule
     & exp. & M1 & M2 & \multicolumn{4}{c}{P\textsubscript{D1} -P\textsubscript{D2}} \\
     & & & & \multicolumn{2}{c}{M1 monomers} & \multicolumn{2}{c}{M2 monomers} \\
     \midrule
     solv.    & 1.82 & 1.81 & 1.84 & 1.78 & 1.81 & 1.80 & 1.84  \\
     no solv. & 1.94 & 1.98 & 1.99 & 1.93 & 1.95 & 1.94 & 1.96  \\
     diff.    & 0.12 & 0.17 & 0.15 & 0.15 & 0.14 & 0.14 & 0.12  \\
     & & \multicolumn{4}{c}{Chl\textsubscript{D1}-P\textsubscript{D1} -P\textsubscript{D2}-Chl\textsubscript{D2} (M1 monomers)} & &  \\
     \midrule
     solv.    &   & 1.76 & 1.78 & 1.81 & 1.84 & & \\
     no solv. &   & 1.90 & 1.92 & 1.95 & 2.00 & & \\
     diff.    &   & 0.14 & 0.14 & 0.14 & 0.16 & & \\
     \bottomrule
     \end{tabular}
     \caption{$Q_y$ excitation for different Chla monomers and dimers calculated using TD-DFT@CAMY-B3LYP/TZP with and without implicit solvation. All values are in eV.}
     \label{tab:environment}
 \end{table*}

In this appendix, we quantify the electrochromatic shift of the excitation energies of two monomeric and dimeric as well as one tetrameric model of the PSII RC due to solvent effects and protein environment using a polarizable continuum model. The $Q_y$ excitation energies calculated using CAMY-B3LYP-TD-DFT/TZP with and without implicit solvation are shown in table~\ref{tab:environment}. Our calculated electrochromatic shifts agree well with experimental values of about 0.12 eV.\cite{Sirohiwal2021} The fact that we are able to reproduce these shifts reliably with a continuum model is surprising since it's physical origin is routed in the asymmetry of the protein matrix. For the low-lying VEEs, the shifts are more or less independent of the employed model system and they are transferable to the other multichromophoric complexes as well.

\section{\label{app:working-eq}Calculating the BSE Hamiltonian}

The most time-consuming step in the solution of the BSE is to build the matrix elements of the 2-particle Hamiltonian, eq. \eqref{davidson-matrix}. Let us denote with the matrix $\mathbf{K}^{(\pm)}$, a column of $\mathbf{A} \pm \mathbf{B}$ as defined in \eqref{davidson-matrix}, in the primary basis. 

Within the density fitting method, we expand products of atomic orbitals in a basis of auxiliary functions. To introduce the PADF variant of this technique, we label atomic orbitals as $\mu, \nu, \kappa, \lambda$, auxiliary functions as $\alpha, \beta, \gamma, \delta$ and atomic centers as $A, B, C \dots$. We also define the convention that $\mu, \alpha \in A$, $\nu, \beta \in B$, $\kappa, \gamma \in C$ and $\lambda, \delta \in D$, i.e. $\mu$ and $\alpha$ are only labelling functions centered on atom $A$, and so on. The PADF expansion of the products of AOs can then be written as
\begin{equation}
\label{fit}
    \chi_{\mu}(\br)\chi_{\nu}(\br) = 
    \begin{cases}
    \displaystyle{
    \sum_{\beta \in B} c_{\mu \nu, \beta} f_{\beta}(\br) +
    \sum_{\alpha \in A} c_{\nu \mu, \alpha} f_{\alpha}(\br)} & A \neq B \\
    \displaystyle{\sum_{\alpha \in A} \frac{1}{2}\left(c_{\nu \mu, \alpha} + c_{\mu \nu, \alpha} \right) f_{\alpha}(\br)} & A = B \;,
    \end{cases} 
\end{equation}
where the factor of $1/2$ in case $A = B$ is introduced to facilitate evaluation with the same algorithm while avoiding double counting. Let us write \eqref{davidson-matrix} in the primary basis as
\begin{equation}
    K^{(\pm)}_{\mu \nu} = - \sum_{\kappa \lambda} b_{\kappa \lambda}W(\omega = 0)_{\mu \kappa \nu \lambda}
    \pm W(\omega = 0)_{\nu \kappa \mu \lambda} \;,
\end{equation}
where the $b_{\kappa\lambda}$ are elements of the transition density matrix and the $K^{(\pm)}_{\mu\nu}$ denote the matrix elements of a column of $(\mathbf{A} \pm \mathbf{B})^{(n + 1)}$.
Inserting \eqref{fit}, the contribution to $\mathbf{K}^{(\pm)}$ for all atom pairs $(A,B)$ is
\begin{equation}
\label{sigmaAll}
\mathbf{K}^{(\pm)^{\mathrm{AB}}} = 
\mathbf{K}^{(\pm)^{\mathrm{AB,I}}} +
\mathbf{K}^{(\pm)^{\mathrm{AB,II}}} +
\mathbf{K}^{(\pm)^{\mathrm{AB,III}}} +
\mathbf{K}^{(\pm)^{\mathrm{AB,IV}}} \;,
\end{equation}
where
\begin{equation}
\begin{aligned}
    \mathbf{K}^{(+)^{\mathrm{AB,III}}} = & \left[\mathbf{K}^{(+)^{\mathrm{AB,II}}}\right]^T \\
    \mathbf{K}^{(-)^{\mathrm{AB,III}}} = & -\left[\mathbf{K}^{(-)^{\mathrm{AB,II}}}\right]^T \;. 
\end{aligned}
\end{equation}
In these and in the following quantities the matrices are restricted to the primary basis functions centered on the atoms denoted by the indices in the superscripts. We define the intermediates
\begin{equation}
\label{intermediateI}
I^{ABC}_{\mu \nu \gamma} = c^{ABB}_{\mu \nu \beta} W(\omega=0)^{BC}_{\beta\gamma} \;,
\end{equation}
and 
\begin{equation}
F^{BAA}_{\nu \mu \alpha} = 
\sum_{\lambda}
b^{DB}_{\lambda \nu}
c^{DAA}_{\lambda \mu  \alpha} \;.
\end{equation}
Here $W(\omega=0)_{\beta\gamma}$ are the matrix elements of the statically screened interaction in the basis of the auxiliary functions $\left\{f_{\alpha}\right\}_{\alpha = 1, \dots, N_{aux}}$,
\begin{equation}
    W(\omega=0)_{\alpha\beta} = \int d\br \int d\br' f_{\alpha}(\br) W(\br,\br',\omega=0) f_{\beta}(\br') \;.
\end{equation}
We can then write
\begin{equation}
\begin{aligned}
K^{\pm, AC,I}_{\mu \kappa} = &
\sum_{\nu \lambda} \sum_{\alpha \gamma}
b^{DB}_{\lambda \nu}
c^{DAA}_{\lambda \mu  \alpha}
W(\omega=0)^{AC}_{\alpha \gamma}  
c^{BCC}_{\nu \kappa  \gamma} \\
& = \sum_{\nu \alpha}  
F^{BAA}_{\nu \mu \alpha} 
I^{BCA}_{ \nu \kappa \alpha,\tau} \\
K^{\pm, AC,II}_{\mu \kappa} = &
\sum_{\nu \lambda} \sum_{\alpha \beta}
b^{DB}_{\lambda \nu}
c^{DAA}_{\lambda \mu  \alpha}
W(\omega=0)^{AB}_{\alpha \beta}  
c^{CBB}_{\kappa  \nu \beta} \\
& = \sum_{\nu \alpha}  
F^{BAA}_{\nu \mu \alpha}
 I^{CBA}_{\kappa \nu \alpha} \\
K^{\pm,AC,IV}_{\mu \kappa} = &
\sum_{\nu \lambda} \sum_{\delta \beta}
b^{DB}_{\lambda \nu}
c^{ADD}_{\mu \lambda  \delta}
W(\omega=0)^{DB}_{\delta \beta}  
c^{CBB}_{\kappa  \nu \beta} \\
= & \sum_{\lambda\delta}  
\sum_{\nu }
b^{DB}_{\lambda \nu}
I^{CBD}_{\kappa \nu \delta}
b^{ADD}_{\mu \lambda  \delta} \;,
\end{aligned}
\end{equation}
where in the $+$ case $b$ is symmetric, and antisymmetric otherwise. These are the working equations with which \eqref{davidson-matrix} is implemented. They are similar to the ones for the self-energy, outlined in ref.~\citen{Forster2020}.

\section{\label{app:lin-dependencies}Elimination of diffuse functions from the primary basis}

In addition to the usual canonical orthonormalization\cite{Lowdin1967} during the SCF prior to the qsGW calculation we herein introduce an additional step in order to improve the numerical stability of our algorithm. To project out too diffuse functions from the primary basis we first diagonalize the overlap matrix of primary basis functions $\mathbf{S}$,
\begin{equation}
    \mathbf{S} = \mathbf{U}^{T} \Lambda\mathbf{U} \;.
\end{equation}
We then remove a column $u_i$ from the transformation matrix if the corresponding eigenvalue $\lambda_i$ is smaller than some predefined threshold $\epsilon_s$. We then define 
\begin{equation}
    \mathbf{V} = \mathbf{U}\mathbf{U}^{T} \;,
\end{equation}
and use this projector to transform all matrices in the primary basis, the Green's functions, the self-energy contributions as well as the matrices defined in \eqref{davidson-matrix} according to
\begin{equation}
    \mathbf{K} = \mathbf{V}^{T}\mathbf{K}'\mathbf{V} \;,
\end{equation}
where $\mathbf{K}'$ would be the original exchange-like matrix in the primary basis including the diffuse part. This transformation is not necessary if a very large auxiliary basis set is used and is switched off in that case. 


\begin{acknowledgement}
Insightful discussions and comments by Leeor Kronik and Souloke Sen are gratefully acknowledged. This research received funding (project number 731.017.417) from the Netherlands Organisation for Scientific Research (NWO) in the framework of the Innovation Fund for Chemistry and from the Ministry of Economic Affairs in the framework of the \enquote{\emph{TKI/PPS-Toeslagregeling}}. 
\end{acknowledgement}


\begin{suppinfo}
All structures used in this work. Additional excitation energies for different geometries. Characterization of excitation energies.
\end{suppinfo}


\bibliography{all}



\section*{Supplementary material}

# Supporting information to: Quasiparticle Self-Consistent $GW$-Bethe-Salpeter equation calculations for large chromphoric systems

Arno Förster* and Lucas Visscher

*Theoretical Chemistry, Vrije Universiteit, De Boelelaan 1083, NL-1081 HV, Amsterdam, The Netherlands*

E-mail: a.t.l.foerster@vu.nl

# 1 VEEs of Chlorophyll dimers for different optimized geometries

Table 1: The lowest six excitation energies of a Chla dimer (monomer geomtry of figure 2a in the main text) All values are in eV. The structures have been optimized in this work at CAM-B3LYP-D3(BJ)/TZP.

| kernel | $\Omega_1$ | $\Omega_2$ | $\Omega_3$ | $\Omega_4$ | $\Omega_5$ | $\Omega_6$ |
|---|---|---|---|---|---|---|
| exp. (VEE)[1] | | | 1.95 (estimated) | | | |
| exp. (band max)[1] | | | 1.90 | | | |
| Ma dimer (figure 2a in main text, 108 atoms) | | | | | | |
| ev$GW$@LDA | 1.98 | 1.99 | 2.16 | 2.22 | 2.51 | 2.64 |
| ev$GW$@PBEH40 | 1.97 | 2.02 | 2.24 | 2.27 | 2.58 | 2.67 |
| qs$GW$ | 1.94 | 1.98 | 2.25 | 2.28 | 2.56 | 2.68 |
| CAMY-B3LYP | 2.12 | 2.16 | 2.38 | 2.43 | 2.51 | 2.61 |
| $\omega$B97-X | 2.05 | 2.10 | 2.63 | 2.68 | 3.10 | 3.27 |
| M2 dimer (figure 2b in main text, 140 atoms) | | | | | | |
| ev$GW$@LDA | 1.96 | 2.00 | 2.17 | 2.24 | 2.48 | 2.64 |
| ev$GW$@PBEH40 | 1.97 | 1.98 | 2.26 | 2.29 | 2.50 | 2.67 |
| qs$GW$ | 1.94 | 1.96 | 2.25 | 2.28 | 2.51 | 2.68 |
| CAMY-B3LYP | 2.12 | 2.14 | 2.34 | 2.42 | 2.49 | 2.61 |
| $\omega$B97-X | 2.06 | 2.08 | 2.65 | 2.67 | 3.04 | 3.27 |
| M3 dimer (figure 2b in main text, 178 atoms) | | | | | | |
| ev$GW$@LDA | 1.98 | 2.02 | 2.08 | 2.11 | 2.31 | 2.42 |
| ev$GW$@PBEH40 | 1.96 | 1.98 | 2.13 | 2.15 | 2.33 | 2.43 |
| qs$GW$ | 1.95 | 1.97 | 2.14 | 2.16 | 2.35 | 2.43 |
| CAMY-B3LYP | 2.15 | 2.18 | 2.25 | 2.32 | 2.38 | 2.43 |
| $\omega$B97-X | 2.10 | 2.11 | 2.57 | 2.61 | 2.84 | 2.90 |

Table 2: The lowest 2 excitations of the Chlorophyll dimer (M3 structure in figure 2b in the main text) optimized at different geometries calculated with different methods.

|  | CAM-B3LYP-D3(BJ) TZP | | CAM-B3LYP-D3(BJ) TZ3P | | B3LYP-D3(BJ) | | PBE-D4 TZP | | PBE | |
|---|---|---|---|---|---|---|---|---|---|---|
| qs$GW$ | 1.94 | 1.98 | 1.92 | 1.96 | 1.82 | 1.88 | 1.83 | 1.85 | 1.84 | 1.86 |
| ev$GW$@LDA | 1.98 | 1.99 | 1.98 | 1.99 | | | 1.86 | 1.88 | | |
| ev$GW$@PBEH40 | 1.97 | 2.02 | 2.00 | 2.04 | | | 1.86 | 1.88 | | |
| CAMY-B3LYP$^a$ | | | | | | | | | 2.03 | 2.08 |
| CAMY-B3LYP$^b$ | 2.13 | 2.16 | | | 1.96 | 2.04 | 2.01 | 2.02 | 2.01 | 2.05 |
| $\omega$B97-X | 2.05 | 2.10 | | | | | | | | |

# 2 VEEs of Chlorophyll dimers for different crystal structures

Table 3: Comparison of the $Q_y$ excitation energies obtained with different methods and experimental values. The geometries are based on chrystal structures. All values are in eV.

|  | D140 | | D164 | |
|---|---|---|---|---|
| ev$GW$@LDA | 1.78 | 1.81 | 1.78 | 1.86 |
| ev$GW$@PBEH40 | 1.71 | 1.75 | 1.73 | 1.77 |
| qs$GW$ | 1.71 | 1.74 | 1.74 | 1.77 |
| CAMY-B3LYP | 1.93 | 1.95 | 1.94 | 1.96 |
| exp. (VEE)[1] | 1.95 (estimated) | | | |
| exp. (band max)[1] | 1.90 | | | |

In contrast to the $GW$-BSE VEEs, the CAMY-B3LYP-TD-DFT results for the crystal structures are in excellent agreement with the available experimental gas-phase data.[2–4] In light of the factors just discussed, the excellent agreement of the CAMY-B3LYP-TD-DFT calculations is most likely due to an overestimation of the true VEEs (compares to the results shown in the main text and in table 2) which then cancels with the errors due to inadequate geometries.

# 3  ev$GW$ single-particle energies of the hexameric complex

The *evGW*@PBEH40 single-particle energies for the hexameric complex shown in table 4 do not change their order compared to the KS-DFT single-particle energies.

Table 4: The five highest occupied and the five lowest unoccpied single-particle energies at the KS-DFT (PBEH40) and the ev*GW*@PBEH40 level of theory. The difference between the energy levels is shown in the last column.

| index | E(KS) [eV] | E(ev$GW$) [eV] | $\Delta_{KS-evGW}$ |
|---|---|---|---|
| | | occupied | |
| 932 | -6.759 | -6.911 | 0.152 |
| 933 | -6.716 | -6.794 | 0.078 |
| 934 | -6.674 | -6.763 | 0.089 |
| 935 | -6.626 | -6.650 | 0.024 |
| 936 | -6.595 | -6.624 | 0.028 |
| | | virtual | |
| 937 | -3.601 | -2.453 | -1.148 |
| 938 | -3.543 | -2.376 | -1.167 |
| 939 | -3.517 | -2.418 | -1.099 |
| 940 | -3.514 | -2.327 | -1.186 |
| 941 | -3.511 | -2.334 | -1.177 |

## 3.1  TD-DFT/$\omega$B97-X/TZP

Table 5: The lowest TD-DFT/$\omega$B97-X/TZP excited states of the hexameric chromophore complex in the RC of PSII.[a]

|  | VEE | $f$ | Character | weight |
|---|---|---|---|---|
| $\Omega_1$ | 1.92 | 0.33 | Chl$_{D2}$* | 0.47 |
| $\Omega_2$ | 1.93 | 0.64 | Pd$_{D2}$* | 0.23 |
|  |  |  | Pd$_{D1}$* | 0.14 |
|  |  |  | Pd$_{D1}^+$ - Pd$_{D2}^-$ | 0.14 |
|  |  |  | Pd$_{D2}^+$ - Pd$_{D1}^-$ | 0.12 |
| $\Omega_3$ | 1.94 | 0.14 | Pd$_{D1}$* | 0.23 |
|  |  |  | Chl$_{D1}$*/Chl$_{D1}^+$ - Pheo$_{D1}^-$ | 0.18 |
|  |  |  | Chl$_{D2}$* | 0.09 |
|  |  |  | Chl$_{D1}$*/Chl$_{D1}^+$ - Pheo$_{D1}^-$ | 0.09 |
| $\Omega_4$ | 1.96 | 0.18 | Pheo$_{D1}$*/Pheo$_{D1}^+$ - Chl$_{D1}^-$ | 0.16 |
|  |  |  | Pheo$_{D1}$*/Pheo$_{D1}^+$ - Chl$_{D1}^-$ | 0.14 |
|  |  |  | Pheo$_{D2}$* | 0.13 |
|  |  |  | Pd$_{D2}$* | 0.09 |
|  |  |  | Pd$_{D1}$* | 0.09 |
| $\Omega_5$ | 1.97 | 0.09 | Pheo$_{D2}$* | 0.34 |
|  |  |  | Chl$_{D2}$* | 0.11 |
| $\Omega_6$ | 1.98 | 0.07 | Chl$_{D1}$* | 0.22 |
|  |  |  | Chl$_{D1}$*/Chl$_{D1}^+$ - Pheo$_{D1}^-$ | 0.17 |
|  |  |  | Pheo$_{D1}$*/Pheo$_{D1}^+$ - Chl$_{D1}^-$ | 0.10 |
|  |  |  | Pheo$_{D1}$* | 0.07 |

[a]Shown are the excitation energies $\Omega_S$ (in eV), the dominant coefficients of the corresponding eigenvector and the associated particle-hole transitions, as well as the oscillator strengths $f$.

## 3.2 qs$GW$@-BSE/TZP

Table 6: The lowest ev$GW$@PBEH40-BSE/TZP excited states of the hexameric chromophore complex in the RC of PSII.[a].

|  | VEE | $f$ | Character | weight |
|---|---|---|---|---|
| $\Omega_1$ | 1.93 | 0.56 | Chl$_{D2}$* | 0.33 |
|  |  |  | Pd$_{D1}$* | 0.32 |
| $\Omega_2$ | 1.94 | 0.48 | Pd$_{D2}$* | 0.52 |
| $\Omega_3$ | 1.96 | 0.10 | Pd$_{D1}$* | 0.28 |
|  |  |  | Pheo$_{D2}$* | 0.24 |
|  |  |  | Chl$_{D2}$* | 0.13 |
| $\Omega_4$ | 1.97 | 0.38 | Pheo$_{D1}$*/Pheo$_{D1}^+$ - Chl$_{D1}^-$ | 0.25 |
|  |  |  | Pheo$_{D1}$*/Pheo$_{D1}^+$ - Chl$_{D1}^-$ | 0.17 |
|  |  |  | Pheo$_{D2}$* | 0.13 |
|  |  |  | Chl$_{D1}$* | 0.12 |
| $\Omega_5$ | 1.98 | 0.08 | Chl$_{D2}$* | 0.26 |
|  |  |  | Pheo$_{D2}$* | 0.20 |
| $\Omega_6$ | 2.00 | 0.11 | Chl$_{D1}$*/Chl$_{D1}^+$ - Pheo$_{D1}^-$ | 0.36 |
|  |  |  | Chl$_{D1}$* | 0.22 |

[a]Shown are the excitation energies $\Omega_S$ (in eV), the dominant coefficients of the corresponding eigenvector and the associated particle-hole transitions, as well as the oscillator strengths $f$.

Table 7: The lowest qs$GW$@-BSE/TZP excited states of the hexameric chromophore complex in the RC of PSII.[a]

|  | VEE | $f$ | Character | weight |
|---|---|---|---|---|
| $\Omega_1$ | 1.89 | 0.22 | $Pd_{D2}^*$ | 0.39 |
|  |  |  | $Chl_{D2}^*$ | 0.22 |
| $\Omega_3$ | 1.90 | 0.77 | $Pd_{D2}^*$ | 0.24 |
|  |  |  | $Pd_{D1}^*$ | 0.14 |
|  |  |  | $Pheo_{D2}^*$ | 0.09 |
|  |  |  | $Pd_{D1}^+ - Pd_{D2}^-$ | 0.09 |
| $\Omega_3$ | 1.91 | 0.04 | $Chl_{D1}^*$ | 0.30 |
|  |  |  | $Pd_{D1}^*$ | 0.24 |
|  |  |  | $Chl_{D1}^+ - Pheo_{D1}^-$ | 0.08 |
| $\Omega_4$ | 1.92 | 0.22 | $Pheo_{D2}^*$ | 0.39 |
|  |  |  | $Chl_{D2}^*$ | 0.16 |
|  |  |  | $Pheo_{D2}^*$ | 0.12 |
|  |  |  | $Chl_{D1}^*$ | 0.09 |
| $\Omega_5$ | 1.94 | 0.01 | $Chl_{D1}^*$ | 0.23 |
|  |  |  | $Chl_{D2}^*$ | 0.18 |
|  |  |  | $Pd_{D1}^*$ | 0.16 |
|  |  |  | $Pd_{D2}^*$ | 0.15 |
| $\Omega_6$ | 1.97 | 0.20 | $Pheo_{D1}^*$ | 0.54 |
|  |  |  | $Pheo_{D1}^- - Chl_{D1}^+$ | 0.21 |
| $\Omega_{13}$ | 2.71 | 0.00 | $Pd_{D2}^+ - Chl_{D2}^-$ | 0.81 |
|  |  |  | $Pd_{D1}^+ - Chl_{D2}^-$ | 0.13 |
| $\Omega_{14}$ | 2.73 | 0.00 | $Pd_{D1}^+ - Chl_{D1}^-$ | 0.70 |
|  |  |  | $Pd_{D1}^+ - Pheo_{D1}^-$ | 0.20 |

[a] Shown are the excitation energies $\Omega_S$ (in eV), the dominant coefficients of the corresponding eigenvector and the associated particle-hole transitions, as well as the oscillator strengths $f$.

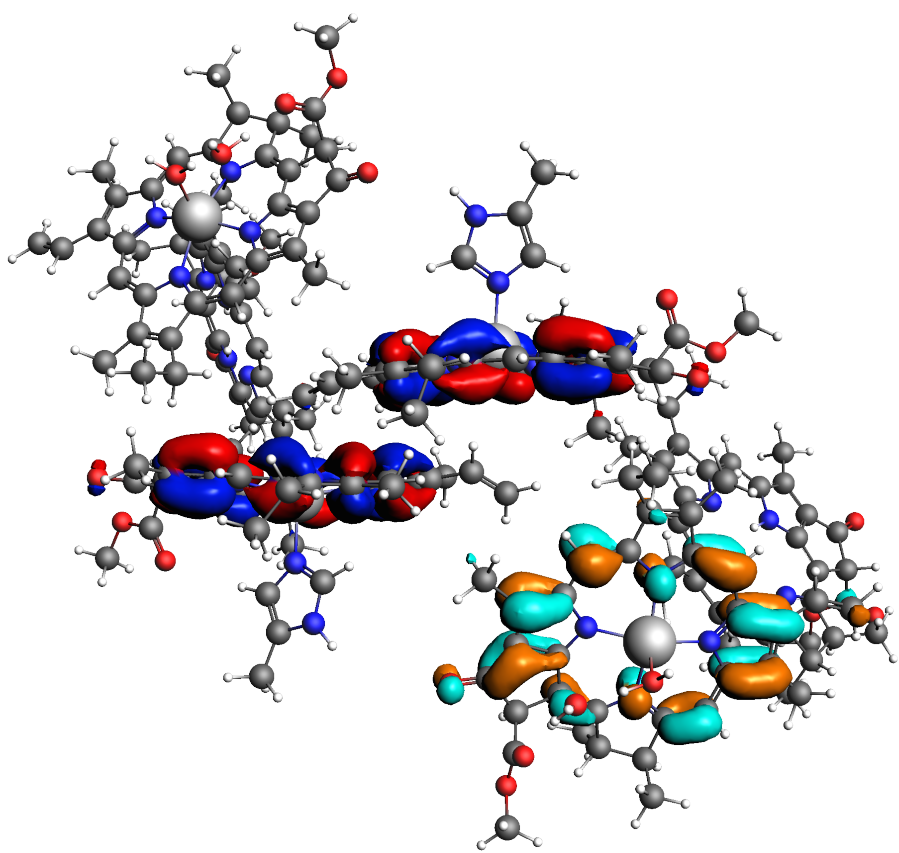

Figure 1: First excited state of the hexameric complex with pronounced CT character using qs$GW$@-BSE/TZP.

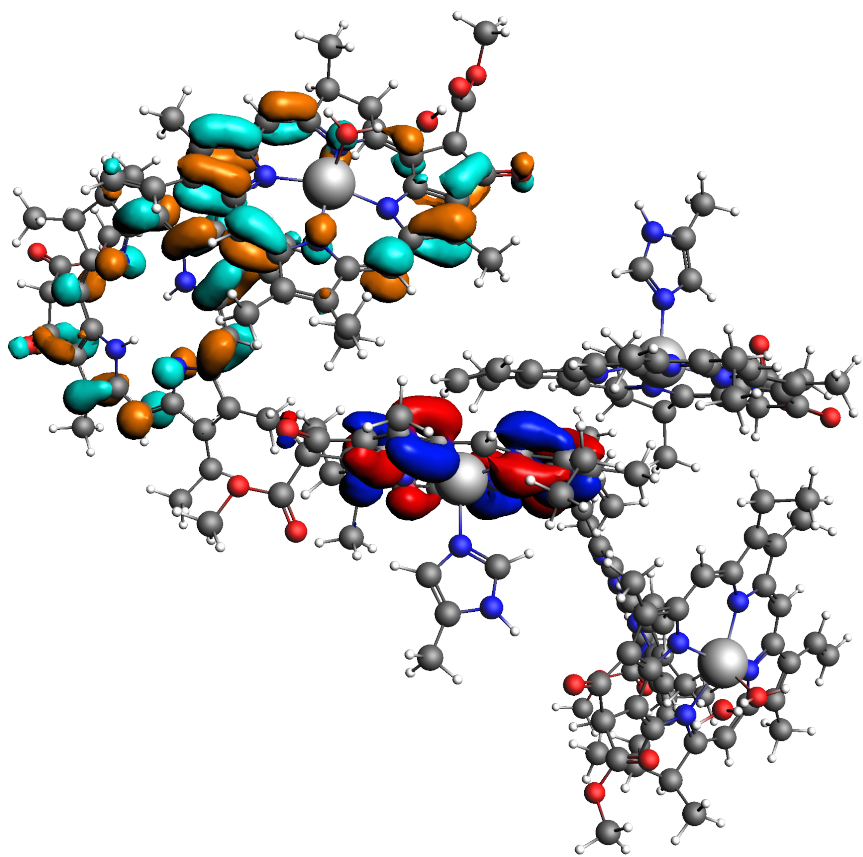

Figure 2: Second excited state of the hexameric complex with pronounced CT character using qs$GW$@-BSE/TZP.




\end{document}